\documentclass{aa}  

\usepackage{graphicx}
\usepackage[varg]{txfonts}
\usepackage{multirow}
\usepackage{amsmath}
\usepackage{lscape}
\usepackage{bm}
\usepackage{ulem}

\usepackage{color}

\usepackage{natbib}
\usepackage{subfigure}
\usepackage{overpic}

\newcommand{\myemail}{hanae.inami@univ-lyon1.fr}

\usepackage{xspace}
\newcommand{\um}{$\,\mu$m\xspace}

\begin{document}

\title{The AKARI $2.5-5$ Micron Spectra of Luminous
  Infrared Galaxies in the Local Universe}

\author{
      H. Inami\inst{1} \thanks{\email{\myemail}}, 
      L. Armus\inst{2}, 
      H. Matsuhara\inst{3},
      V. Charmandaris\inst{4,5},
      T. D\'iaz-Santos\inst{6},
      J. Surace\inst{2},
      S. Stierwalt\inst{7,8},
      Y. Ohyama\inst{9},
      J. Howell\inst{2},
      J. Marshall\inst{10},
      A. S. Evans\inst{11,12},
      S. T. Linden\inst{11},
      J. Mazzarella\inst{2}
}

\institute{ 
  Univ Lyon, Univ Lyon1, Ens de Lyon, CNRS, Centre de Recherche
  Astrophysique de Lyon (CRAL) UMR5574, F-69230, Saint-Genis-Laval,
  France
  \and 
  Infrared Processing and Analysis Center, Caltech, 1200
  E. California Blvd., Pasadena, CA 91125, USA
  \and 
  Institute of Space and Astronautical Science, Japan Aerospace
  Exploration Agency, Japan
  \and 
  Institute for Astronomy, Astrophysics, Space Applications \& Remote
  Sensing, National Observatory of Athens, GR-15236, Penteli, Greece
  \and 
  University of Crete, Department of Physics, GR-71003, Heraklion,
  Greece
  \and 
  N\'ucleo de Astronom\'ia de la Facultad de Ingenier\'ia, Universidad
  Diego Portales, Av. Ej\'ercito Libertador 441, Santiago, Chile 
  \and 
  Department of Astronomy, University of Virginia, P.O. Box 400325,
  Charlottesville, VA 22904, USA
  \and 
  National Radio Astronomy Observatory, 520 Edgemont Road,
  Charlottesville, VA 22903, USA
  \and 
  Academia Sinica, Institute of Astronomy and Astrophysics, 11F of
  Astronomy-Mathematics Building, National Taiwan University, No.1,
  Sec. 4, Roosevelt Rd, 10617 Taipei, Taiwan R.O.C.
  \and 
  Glendale Community College, 1500 North Verdugo Road, Glendale, CA
  91208, USA
  \and 
  Department of Astronomy, University of Virginia, P.O. Box 400325,
  Charlottesville, VA 22904, USA
  \and
  National Radio Astronomy Observatory, 520 Edgemont Road,
  Charlottesville, VA 22903, USA
}

\date{Updated \today}

\abstract { We present {\it AKARI} $2.5-5$\um spectra of 145 local
  luminous infrared galaxies (LIRGs; $L_{IR} \geq 10^{11} \, L_\odot$)
  in the Great Observatories All-sky LIRG Survey (GOALS).  In all of
  the spectra, we measure the line fluxes and equivalent widths (EQWs)
  of the polycyclic aromatic hydrocarbon (PAH) at 3.3\um and the
  hydrogen recombination line Br$\alpha$ at 4.05\um, with apertures
  matched to the slit sizes of the {\it Spitzer} low-resolution
  spectrograph and with an aperture covering $\sim 95\%$ of the total
  flux in the {\it AKARI} 2D spectra.  The star formation rates (SFRs)
  derived from the Br$\alpha$ emission measured in the latter aperture
  agree well with SFRs estimated from $L_{IR}$, when the dust
  extinction correction is adopted based on the 9.7\um silicate
  absorption feature.  Together with the {\it Spitzer}/IRS $5.2-38$\um
  spectra, we are able to compare the emission of the PAH features
  detected at 3.3\um and 6.2\um.  These are the two most commonly used
  near/mid-infrared indicators of starburst or active galactic nucleus
  (AGN) dominated galaxies.  We find that the 3.3\um and 6.2\um PAH
  EQWs do not follow a linear correlation and at least a third of the
  galaxies classified as AGN-dominated sources using the 3.3\um
  feature are classified as starbursts based on the 6.2\um feature.
  These galaxies have a bluer continuum slope than galaxies that are
  indicated to be starburst-dominated by both PAH features.  The bluer
  continuum emission suggests that their continuum is dominated by
  stellar emission rather than hot dust.  We also find that the median
  {\it Spitzer}/IRS spectra of these sources are remarkably similar to
  the pure starburst-dominated sources indicated by high PAH EQWs in
  both 3.3\um and 6.2\um.  Based on these results, we propose a
  revised starburst/AGN diagnostic diagram using $2-5$\um data: the
  3.3\um PAH EQW and the continuum color,
  $F_\nu(4.3{\rm \mu m})/F_\nu(2.8{\rm \mu m})$.  We use the {\it
    AKARI} and {\it Spitzer} spectra to examine the performance of our
  new starburst/AGN diagnostics and to estimate 3.3\um PAH fluxes
  using the {\it JWST} photometric bands in the redshift range
  $0 < z < 5$.  Of the known PAH features and mid-infrared high
  ionization emission lines used as starburst/AGN indicators, only the
  3.3\um PAH feature is observable with {\it JWST} at $z > 3.5$,
  because the rest of the features at longer wavelengths fall outside
  the {\it JWST} wavelength coverage.  }

\keywords{Galaxies: starburst -- Galaxies: active -- Infrared: galaxies}

\titlerunning{AKARI $2.5-5$\um Spectra of Local LIRGs in GOALS}
\authorrunning{Inami et al.}

\maketitle



\section{Introduction}\label{sec:intro}

The rest-frame $2-5$\um spectra of galaxies provide a wealth of
diagnostic power, since they can be used to trace hot dust, small
grain dust emission, ionizing flux, and starlight.
Starburst-dominated galaxies show a strong polycyclic aromatic
hydrocarbon (PAH) emission feature at 3.3\,\um
\citep[e.g.,][]{Genz00,Iman10}, whereas galaxies with an obscured
active galactic nucleus (AGN) show a rising (power law) continuum
\citep[e.g.,][]{Armu07,YamaR13}.  The Br$\alpha$ hydrogen
recombination line at $4.05$\um provides a direct measure of the
ionizing radiation and thus star formation rate. The absorption
features of $\rm H_2O$ ($3.05$\um) ice, $\rm CO_2$ ($4.27$\um) ice,
and CO ice and gas ($4.67$\um) may be present as well.  The {\it
  AKARI} Infrared Satellite \citep{Mura07} performed an all-sky
imaging survey in the mid- \citep[$9$ and $18$\um,][]{Ishi10} and
far-infrared \citep[$65$, $90$, $140$, and $160$\um,][]{Doi15,Taki15},
but it also had the capability to carry out pointed observations using
the Infrared Camera (IRC) covering $2.5-5$\um at a spectral resolving
power of $R \sim 120$ \citep[IRC;][]{Onak07}.

In infrared luminous galaxies with a dominant central AGN, hot dust
($\sim 1000\,$K) emission can dominate the near-infrared
continuum. The $2.5-5$\um coverage of {\it AKARI} facilitates
measurements of the relative amounts of hot, warm, and cold dust via
comparison to {\it Spitzer} mid- and far-infrared data.  The PAH
emission features are excited by individual UV photons, and therefore
they are a direct probe of star formation \citep[e.g.,][]{Peet04} and
are prevalent in star forming galaxies \citep{Smit07}.  In addition,
the PAH features in starburst-dominated galaxies are an important
diagnostic of the dust grain properties. For example, weak PAH
emission at 3.3\um, relative to other mid-infrared PAH features (e.g.,
$7.7$\um), suggests ionization of the small dust grains
\citep{Drai07a}.  On the other hand, PAH features are invariably
absent from the spectra of AGN, which have very hard radiation fields
and lack Photo-dissociation Regions \citep[PDRs, e.g.,][]{Laur00,
  Weed05}.  For the upcoming James Webb Space Telescope ({\it JWST}),
the 3.3\um PAH emission feature will be the only PAH feature
observable at $z > 3.5$.  It is, therefore, important to characterize
this starburst/AGN diagnostic feature in a large unbiased sample in
the local Universe and to quantitatively tie the diagnostic power of
this feature to the stronger, and more commonly used mid-infrared PAH
features that have been used to probe the physics of active starburst
galaxies for over two decades.

A large number of studies, some of which include local LIRGs, have
explored some of the properties of the $3.3$\um PAH, Br$\alpha$, and
hot dust detected with the {\it AKARI} spectrograph.
For example, these features were used to investigate the starburst/AGN
diagnostics of (U)LIRGs and to discuss the properties of starburst and
AGN \citep{Iman08,Iman10,Lee12}.  \cite{Woo12} focused on Type~I AGN
and found that the AGN activity is directly related to the nuclear
starburst but not the global star formation of the host galaxy.  In
addition, as a potential star formation rate (SFR) indicator, 3.3\um
PAH emission was compared against the total infrared emission
\citep{Kim12,YamaR13,Yano16,Mura17}. The 3.3\um PAH and infrared
luminosities correlate well up to $10^{11}\,L_\odot$, then break down.
However, most of these studies have not directly compared the key
diagnostic features over the entire mid-infrared range by combining
the {\it AKARI} data with those taken with the Infrared Spectrograph
(IRS) on the {\it Spitzer} Space Telescope \citep{Houc04} which
covered the $5-38$\um range, and for which large studies of starbursts
and AGN have been published
\citep[e.g.,][]{Bran06,Gall10,Haan11b,Stie13,Stie14,Ship16}.  In this
paper, we use aperture-matched {\it AKARI} and {\it Spitzer}/IRS
spectra of a sample of 145 local Luminous Infrared Galaxies (LIRGs,
$L_{IR} \geq 10^{11} \, L_\odot$) to further calibrate the $3.3$\um
PAH feature diagnostics, and set the stage for understanding the
spectroscopic and photometric observations of this feature in
high-redshift galaxies with {\it JWST}.

Our 145 local LIRG targets are taken from the Great Observatories
All-sky LIRG Survey \citep[GOALS;][]{Armu09}.  The orignal intention
was to observe the complete GOALS sample with {\it AKARI}, but
unfortunately the mission ended in the middle of the project due to
the depletion of the liquid helium coolant. The entire sample of GOALS
galaxies (nuclei) and the subsample in this work cover the same
$\log(L_{IR}/L_\odot)$ range of $11.00-12.57$, but with mean (median)
values of 11.48 (11.40) and 11.60 (11.61), respectively.  The ranges
of luminosity distances ($D_L$) are $15.9-400.0\,{\rm Mpc}$ and
$17.9-395.0\,{\rm Mpc}$ with mean (median) values of
$115.1\,{\rm Mpc}$ ($95.2\,{\rm Mpc}$) and $132.0\,{\rm Mpc}$
($117.5\,{\rm Mpc}$), respectively.  The GOALS sample in total
contains 178 LIRG and 22 ultra luminous infrared galaxy (ULIRG,
$L_{IR} \geq 10^{12} \, L_\odot$) systems in the local Universe.
These objects are a flux limited sample drawn from the {\it IRAS}
Revised Bright Galaxy Sample \citep[RBGS;][]{Sand03}, which covers
galactic latitudes greater than five degrees and includes 629
extragalactic objects with $60\,\mathrm{\mu m}$ flux densities greater
than 5.24~Jy.  Although LIRGs are not common in the local Universe,
their number density increases rapidly with redshift at least up to
$z \sim 3$ and their star formation rate density dominates at
$z \sim 2-3$, the peak of galaxy formation in the Universe
\citep[e.g.,][]{Mada14}.  Thus, observations of local LIRGs can help
improve our understanding of obscured star formation and black hole
growth at all epochs.  Since the PAH features are often very strong
\citep[e.g.,][]{Smit07}, and easily seen at high redshifts even in
low-resolution spectra when no other features are visible, it is
critically important to assess their diagnostic power across the
entire infrared spectral regime in local galaxies, where high
signal-to-noise, multi-wavelength data can be brought to bear to
understand the underlying heating mechanisms.

In this study, we combine the {\it AKARI} $2.5-5\,{\rm \mu m}$ spectra
and the {\it Spitzer} $5.2-38$\um spectra to investigate the nuclear
energy sources, starburst ages, star formation rates, gas ionization,
and dust properties of local LIRGs. We will also report on possible
diagnostics for both local and high redshift galaxies using {\it JWST}
observations of this important spectral region.  This paper is
organized as follows: the {\it AKARI} observations are described in
Section~\ref{sec:obs}, followed by the data reduction and analysis in
Section~\ref{sec:data}.  The {\it AKARI} spectra and derived
properties with the {\it Spitzer} data are shown in
Section~\ref{sec:results}. We present our interpretation of the
results in Section~\ref{sec:discussion}.  We also briefly discuss how
one could observe these targets with {\it JWST}.  The summary and
conclusions are given in Section~\ref{sec:summary}. A cosmology of
$\Omega_\Lambda = 0.72$, $\Omega_m = 0.28$
$H_0 = 70 \, \mathrm{km \, s^{-1} \,Mpc^{-1}}$ is adopted throughout.


\section{Observations}\label{sec:obs}

The $2.5-5$\um spectra of 145 GOALS sources were observed with {\it
  AKARI} pointed observations between 2006 October 29 and 2010
February 14 using the Infrared Camera \citep[IRC;][]{Onak07}. The data
obtained prior to 2007 August were taken during the cold phase, and
the data from after this time were taken in the warm (post-helium)
phase mission. The total number of pointings was 385 with the
majority obtained with the AGNUL program (54\%, PI T. Nakagawa), in
addition to 36\% from the GOALS program (PI H. Inami), 8\% from DTIRC (PI
H. Murakami), and the remaining 2\% from the BRSFR (PI M. Malkan),
EGANS (PI H. Matsuhara), and NULIZ programs (PI H.-S. Hwang).  The
\textsf{Np} aperture with a size of $1\arcmin \times 1\arcmin$ was
used in order to avoid the confusion of spectra from the entire IRC
field-of-view ($10\arcmin \times 10\arcmin$).  The spectral resolving
power was $R \sim 120$ at $3.5$\um for a point-like source.

Two types of Astronomical Observation Templates (AOTs) were employed
-- IRC04 and IRCZ4. The former was operated during the cold phase
mission without dithering and the latter during the warm phase mission
with dithering. Both of the AOTs obtained data with the same
procedure; taking four spectra, a reference image, and then four or
five (depending on if the satellite maneuver happened during the fifth
frame) spectra at the end of the sequence. The exposure time for each
spectrum was 44.41 seconds.

\section{Data Reduction and Analysis}\label{sec:data}

The standard data reduction software package provided by JAXA, the IRC
Spectroscopy Toolkit for Phase 3 Data\footnote{Available at
  \url{http://www.ir.isas.jaxa.jp/AKARI/Observation/support/IRC/}}
(version 20101025), was used to perform dark subtraction, linearity
correction, flat correction, background subtraction, wavelength
calibration, spectral inclination correction, and spectral response
calibration. This package can handle both the cold and warm phase
data.  Although the number of bad pixels increased dramatically in the
post-helium mission due to the elevated detector temperatures, the
data quality was uniform enough for the purposes of this work, to
measure the 3.3\um PAH and Br$\alpha$ line fluxes and the continuum
emission.  The wavelength accuracy was $\sim0.01$\um and the
uncertainty in absolute flux calibration was estimated to be
$\sim10$\% \citep{Ohya07}.

We extracted the 1D spectrum from the 2D spectral image using two
different methods: (1) the fixed aperture widths of $4.4\arcsec$ (3
pixels) and $10.5\arcsec$ (7 pixels) and (2) an aperture size covering
$\sim95$\% of the total flux in the {\it AKARI} 2D spectra. The 1D
spectra extracted with the former aperture (1) were used for
comparisons with the {\it Spitzer} IRS Short-wavelength Low-resolution
(SL, $5-14.5$\um) and Long-wavelength Low-resolution (LL, $14-38$\um)
spectra, respectively. The $4.4\arcsec$ width is slightly larger than
the SL slit width (3.6\arcsec) but matches well the spatial resolution
of {\it AKARI} over this wavelength range ($\sim 4\arcsec$), providing
an accurate point source spectrum with which to directly compare to
the {\it Spitzer} SL spectra.  At the median distance of our sample,
$117.5 \, {\rm Mpc}$, angular sizes of $4.4\arcsec$ and $10.5\arcsec$
correspond to physical sizes of $2.4 \, {\rm kpc}$ and
$5.7 \, {\rm kpc}$, respectively.  In this work, the LL aperture size
is used only when the full range of {\it AKARI} and {\it Spitzer}
$2.5-38$\um spectra is shown in a figure or needed for an analysis. In
this case, the scaling factor to correct the mismatch between the SL
and LL spectra (due to the different aperture sizes) is applied to the
SL spectrum in the same manner as described in \cite{Stie13}. The
latter aperture (2) was used to obtain spatially-integrated {\it
  AKARI} measurements suitable for investigating integrated properties
of the targets.

The 3.3\um PAH and Br$\alpha$ line fluxes were obtained by fitting a
Gaussian plus a linear function to each feature and its local
continuum (see also Figure~\ref{fig:akari_IRS_spec}). The wavelength
range and function chosen to fit the continuum varied slightly from
source to source to avoid contamination from the broad absorption
features which can be present at $3.05$\um ($\rm H_2O$ ice) and
$4.67$\um (CO ice and gas).  For example, for UGC~05101, a three
degree Chebyshev polynomial had to be employed instead of a linear fit
to the 3.3\um local continuum because its PAH emission feature sits on
the deep absorption features. Typically, the wavelength regions
selected for fitting the continuum were $2.65-2.8$\um (shortward of
the ice feature) and $3.55-3.7$\um for the PAH emission, and
$3.80-4.0$\um and $4.1-4.25$\um for the Br$\alpha$ emission.  When an
emission line was not detected, a $3\sigma$ upper limit was estimated
using a Gaussian function with a height three times the local
continuum dispersion and with a width equal to the {\it AKARI}
intrinsic resolution ($R \sim 120$).

The measured line fluxes and {\it AKARI} spectra of the entire sample
are reported in Table~\ref{tbl:fluxes} and
Appendix~\ref{app:akari_spec}, respectively. Out of the 145 LIRG
sample, 3.3\um PAH emission and Br$\alpha$ emission features were
detected in 133 (92\%) and 91 (63\%) sources, respectively.

\longtab{

  \begin{landscape}

    \begin{center}
      \begin{longtable}{lccccccccc}
        \caption{Line flux and EQW of 3.3\um PAH and Br$\alpha$ and the $F_\nu(4.3{\rm \mu m})/F_\nu(2.8{\rm \mu m})$ color measured in the $4.4\arcsec$ radius aperture}\\
        \hline\hline
        Object name & R.A. & Dec & $F_{\rm 3.3\mu m\,PAH}$ & 
        $\rm EQW_{\rm 3.3\mu m\,PAH}$ & $F_{{\rm Br}\alpha}$ & 
        $\rm EQW_{{\rm Br}\alpha}$ & $F_\nu(4.3{\rm \mu m})/F_\nu(2.8{\rm \mu m})$ & 
        $\rm EQW_{\rm 6.2\mu m\,PAH}$  \\
        \hline
             NGC0023                             &    2.472551 &   25.923980 & $ 7.49 \pm  0.55$ & $ 0.055 \pm  0.004$ & $ 0.60 \pm  0.09$ & $ 0.008 \pm  0.001$ & $ 0.71 \pm  0.02$ & $ 0.579 \pm  0.007$  \\
             NGC0034                             &    2.776792 &  -12.107964 & $ 4.32 \pm  0.23$ & $ 0.046 \pm  0.002$ & $ 0.29 \pm  0.09$ & $ 0.005 \pm  0.002$ & $ 0.93 \pm  0.02$ & $ 0.453 \pm  0.025$  \\
              Arp256                             &    4.711997 &  -10.376844 & $ 4.52 \pm  0.10$ & $ 0.156 \pm  0.003$ & $ 0.58 \pm  0.07$ & $ 0.030 \pm  0.004$ & $ 1.05 \pm  0.07$ & $ 0.719 \pm  0.010$  \\
             NGC0232                             &   10.690820 &  -23.561306 & $ 3.93 \pm  0.22$ & $ 0.048 \pm  0.003$ & $ 0.33 \pm  0.11$ & $ 0.007 \pm  0.002$ & $ 0.85 \pm  0.02$ & $ 0.549 \pm  0.007$  \\
       MCG+12-02-001                             &   13.516243 &   73.084991 & $10.78 \pm  0.25$ & $ 0.134 \pm  0.003$ & $ 1.68 \pm  0.11$ & $ 0.030 \pm  0.002$ & $ 1.29 \pm  0.02$ & $ 0.654 \pm  0.008$  \\
       MCG-03-04-014                             &   17.537453 &  -16.852654 & $ 5.71 \pm  0.12$ & $ 0.115 \pm  0.002$ & $ 0.37 \pm  0.12$ & $ 0.012 \pm  0.004$ & $ 0.98 \pm  0.02$ & $ 0.665 \pm  0.009$  \\
         ESO244-G012                             &   19.534571 &  -44.462093 & $ 6.28 \pm  0.17$ & $ 0.124 \pm  0.003$ & $ 0.79 \pm  0.07$ & $ 0.023 \pm  0.002$ & $ 1.03 \pm  0.02$ & $ 0.665 \pm  0.010$  \\
         CGCG436-030                             &   20.010996 &   14.361864 & $ 2.60 \pm  0.08$ & $ 0.069 \pm  0.002$ & $ 0.53 \pm  0.04$ & $ 0.017 \pm  0.001$ & $ 1.81 \pm  0.05$ & $ 0.347 \pm  0.008$  \\
         ESO353-G020                             &   23.713484 &  -36.137360 & $ 3.12 \pm  0.22$ & $ 0.048 \pm  0.003$ & $ 0.51 \pm  0.00$ & $ 0.013 \pm  0.000$ & $ 0.85 \pm  0.02$ & $ 0.544 \pm  0.008$  \\
               RR032                             &   24.097355 &  -37.321678 & $ 4.12 \pm  0.27$ & $ 0.098 \pm  0.006$ &         $<  0.09$ &          $<  0.003$ & $ 1.02 \pm  0.04$ & $ 0.522 \pm  0.010$  \\
     IRASF01364-1042                             &   24.720614 &  -10.453101 & $ 0.58 \pm  0.08$ & $ 0.044 \pm  0.006$ &         $<  0.10$ &          $<  0.013$ & $ 0.66 \pm  0.10$ & $ 0.392 \pm  0.011$  \\
            IIIZw035  \tablefootmark{$^\dagger$} &   26.126998 &   17.101369 & $ 1.18 \pm  0.06$ & $ 0.062 \pm  0.003$ &         $<  0.13$ &          $<  0.010$ & $ 0.98 \pm  0.05$ &               $...$  \\
        \hline
        \label{tbl:fluxes}
      \end{longtable}
      \tablefoot{ The full table is available at the CDS. The line fluxes and EQWs are in units of
        $10^{-16} \, {\rm W \, m^{-1}}$ and microns, respectively.
        The 6.2\um PAH EQWs (microns) are from \cite{Stie13}.
        \tablefoottext{$\dagger$}{Mismatched pointings between 
        the {\it AKARI} and {\it Spitzer} observations or 
        too extended morphologies to  combine the {\it AKARI} and
        {\it Spitzer} spectra. } Other quantities
        for the GOALS galaxies, such as the AGN fraction (see Section~\ref{ss:paheqws}), 
        stellar mass, UV and IR fluxes, $L_{IR}$, can be found in 
        \cite{Diaz17}, \cite{Howe10}, and \cite{Armu09}. }
   \end{center}

 \end{landscape}

}


\section{Results}\label{sec:results}

\begin{figure}
  \begin{center}
    \includegraphics[width=\textwidth/3,angle=90,trim=-10 0 13 -10,clip]
    {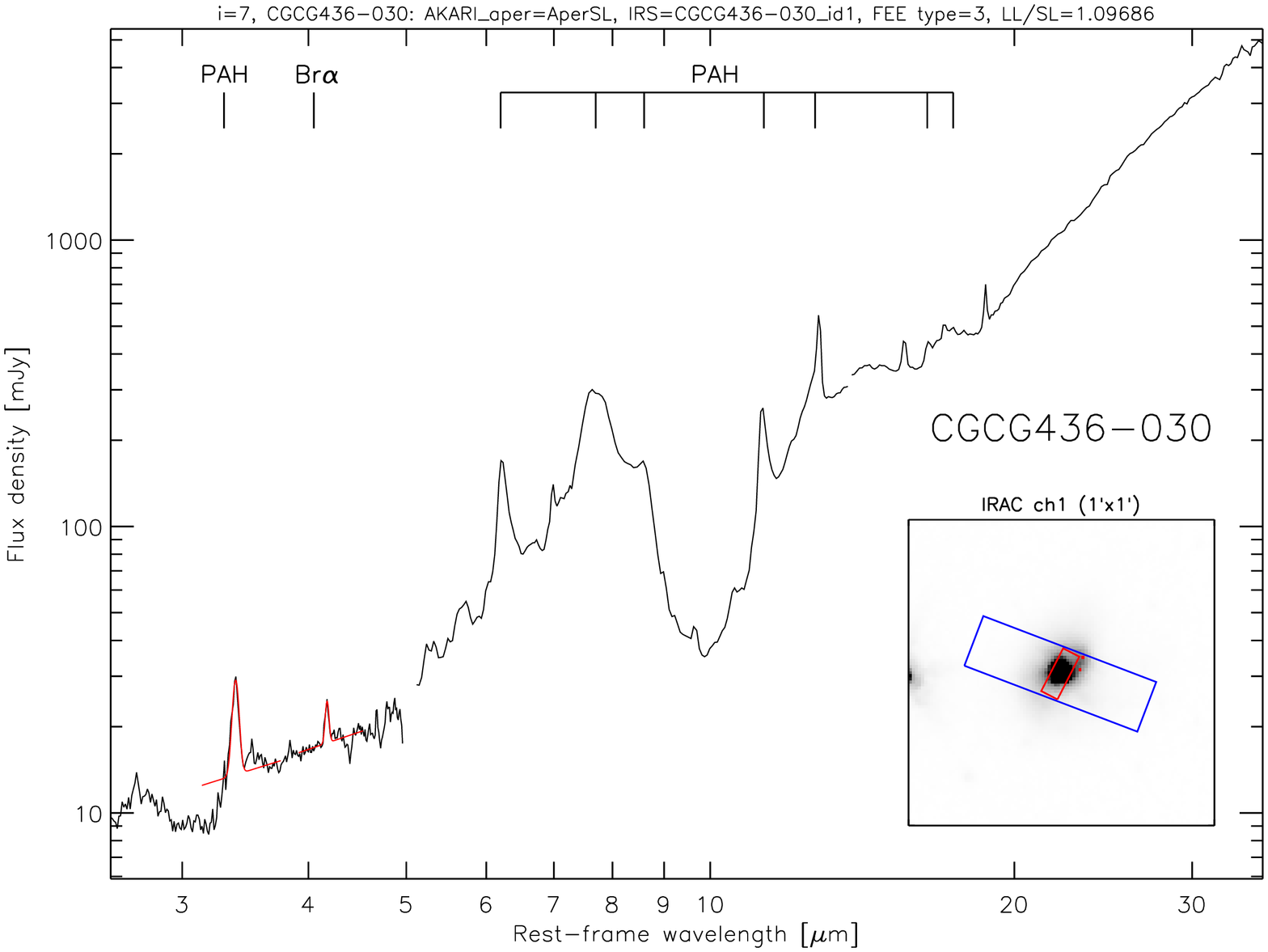}
    \includegraphics[width=\textwidth/3,angle=90,trim=-10 0 13 -10,clip]
    {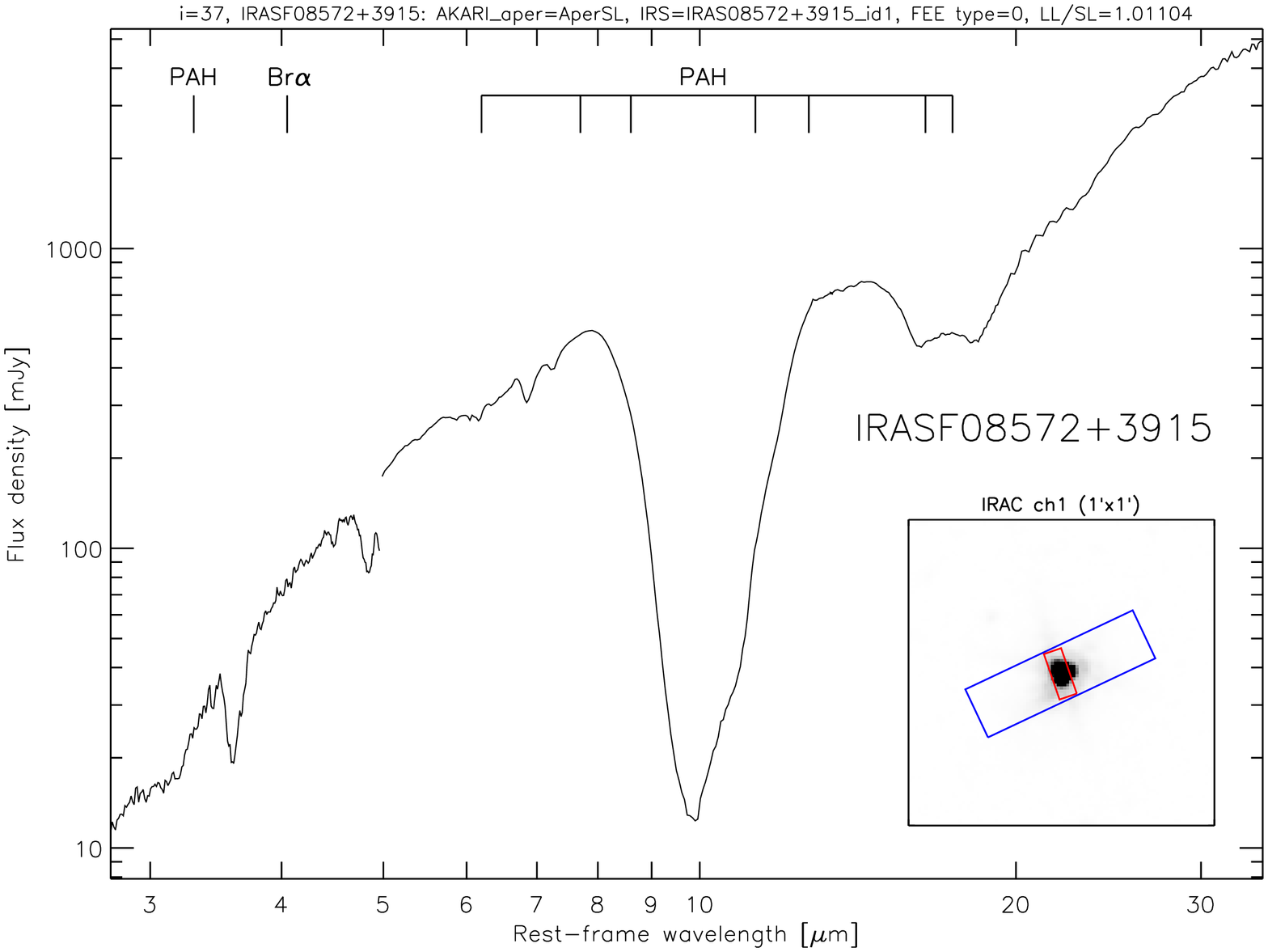}
    \caption{Joined {\it AKARI} and {\it Spitzer} low-resolution
      spectra of two typical LIRGs in the sample. The {\it AKARI}
      spectral extraction is performed with the $10.7\arcsec$ IRS/LL
      aperture size and the IRS/SL spectra are scaled up to match the
      LL spectra (see Section~\ref{sec:data}). A starburst-dominated
      source (CGCG~436-030; top) and an AGN-dominated source
      (IRAS~F08572+3915; bottom) are shown.  Their AGN fractions are
      $8.9 \pm 1.8 \%$ and $46.6 \pm 8.7 \%$, respectively \citep[see
      \S~\ref{ss:paheqws} and][]{Diaz17}. For CGCG~436-030, the fits
      on the 3.3\um PAH feature and the Br$\alpha$ line are denoted
      with red lines.  The IRAC 3.6\um image
      ($1\arcmin \times 1\arcmin$, the same as the {\it AKARI}
      \textsf{Np} FoV size) with the {\it Spitzer} IRS SL (red) and LL
      (blue) apertures overlaid is presented as the inset at the
      bottom right of each panel. The indicated PAH features in the
      {\it Spitzer} wavelength range are at 6.2, 7.7, 8.6, 11.3, and
      12.7\um in SL, and 16.4 and 17.4\um in LL.}
    \label{fig:akari_IRS_spec}
  \end{center}
\end{figure}

Two representative, combined {\it AKARI} and {\it Spitzer}/IRS
low-resolution spectra are shown in
Figure~\ref{fig:akari_IRS_spec}. The one is a starburst-dominated
galaxy (CGCG~436-030; upper panel) and the other is an AGN-dominated
galaxy (IRAS~F08572+3915; lower panel).  As expected, the starburst
CGCG~436-030 shows pronounced PAH emission features, which are absent
in IRAS~F08572+3915. Instead, IRAS~F08572+3915 has a steeply rising
continuum emission contributed by hot and warm dust emission as well
as strong absorption features.

There are 15 sources that show large jumps between extracted {\it
  AKARI} and {\it Spitzer}/IRS spectra.  We exclude these sources from
the discussion of the joint spectral analysis, but include them in the
discussion of the {\it AKARI} spectra alone.

\subsection{Starburst and AGN Classifications Based on PAH Features}\label{ss:paheqws}

The most prominent feature in the $2.5-5$\um spectral range covered by
{\it AKARI} is the emission band at 3.3\um, which arises from C-H
stretching vibrational modes of the PAH carriers.  The moderate
spectral resolution of our data ($R \sim 120$) allows us to separate
the 3.3\um PAH feature from an emission feature at $3.4$\um.  The
exact nature of the $3.4$\um feature is not entirely clear
\citep{Tiel08}, but it is thought to be a PAH satellite feature
\citep{Toku91, Stur00}.  In order to avoid confusion and offer direct
comparisons with past work, we employ only the main 3.3\um PAH emission
throughout this paper.

Because the PAH features are stochastically heated by UV photons and
the mid-infrared continuum can be dominated by hot dust in sources
with powerful AGN, the EQW of the $3.3$\um and $6.2$\um PAH features
have been used as diagnostics of relative AGN power in galactic
nuclei.  For example, \cite{Moor86} and \cite{Iman10} consider that a
source is dominated by an AGN when 3.3\um PAH EQW $<\,0.04$\um.
AGN-dominated galaxies tend to show 6.2\um PAH EQW $<\,0.2$\um
\citep[e.g.,][]{Bran06,Petr11}.  Note that sources which do not
fulfill these criteria are not necessarily starburst-dominated
systems.  That is, galaxies with 6.2\um PAH EQW in the $0.2-0.4$\um
range can be composite sources, while pure starbursts generally show
6.2\um PAH EQW $\gtrsim\,0.6$\um \citep{Bran06, Armu07, Petr11}.

We use the term ``AGN fraction'' to indicate the contribution of AGN
to the total bolometric luminosity of a galaxy, using the multiple AGN
indicators in the mid-infrared from the {\it Spitzer}/IRS data:
[NeV]/[NeII] \citep{Inam13}, [OIV]/[NeII] \citep{Inam13}, 6.2\um PAH
EQW \citep{Stie13}, $S_{30{\rm \mu m}}$/$S_{15{\rm \mu m}}$
\citep{Stie13}, and the Laurent diagram \citep{Petr11,Laur00}.  In
this paper, instead of using the AGN fractions from these individual
AGN indicators, we use the mean of the bolometrically-corrected AGN
fraction of all of them \cite[Table 2 in][and references
therein]{Diaz17}.  The median value of the standard deviation among
these indicators is 3.5\% for our sample.

\begin{figure}
  \begin{center}
    \includegraphics[width=\textwidth/2]
    {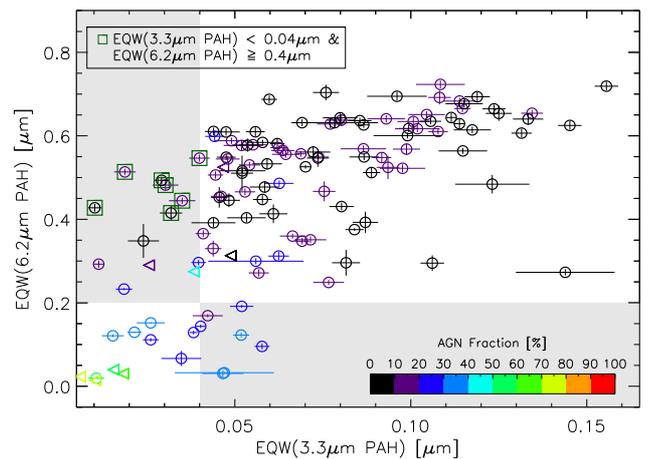}
    \caption{ Comparison between {\it AKARI} 3.3\um PAH EQW and {\it
        Spitzer}/IRS 6.2\um PAH EQW of the GOALS LIRG sample.  The
      circles represent the sources with detections of both PAH
      features, whereas the left-pointing triangles represent
      detections of the 6.2\um PAH but upper limits for the 3.3\um
      PAH.  The symbols are color coded by the AGN bolometric fraction
      in the mid-infrared \citep[Section~\ref{ss:paheqws};][]{Diaz17},
      which is indicated by the color bar shown in the bottom
      right. Galaxies with $\rm PAH(3.3{\rm \mu m}) EQW < 0.04$\um and
      $\rm PAH(6.2{\rm \mu m}) EQW \geq 0.4$\um are also highlighted
      by the green squares.  The gray regions are the areas where the
      implied galactic energy source (starburst or AGN) disagrees
      between 3.3\um PAH EQW and 6.2\um PAH EQW (high 3.3\um PAH EQW
      but low 6.2\um PAH EQW, or vice versa).}
    \label{fig:PAH6_EQW_vs_PAH3_EQW}
  \end{center}
\end{figure}

In Figure~\ref{fig:PAH6_EQW_vs_PAH3_EQW}, we directly compare the
6.2\um PAH EQW and the 3.3\um PAH EQW for the sample.  While there is
a large dispersion in both quantities of
Figure~\ref{fig:PAH6_EQW_vs_PAH3_EQW}, there is, as expected, a broad
correlation of $3.3$ and $6.2$\um PAH EQW among LIRGs.  For most
galaxies, the $6.2$\um PAH EQW asymptotes to the value seen in pure
starburst galaxies ($\sim 0.6-0.7$\um) for $3.3$\um PAH EQW
$> 0.08$\um, although there are notable exceptions.  Except for two
sources with $\rm PAH(3.3\mu m)\,EQW > 0.1$\um, galaxies with
$\rm 0.2\mu m \leq PAH(6.2\mu m)\,EQW < 0.4\mu m$ and
$\rm PAH(6.2\mu m)\, EQW < 0.2$\um have 3.3\um PAH EQWs between
$\sim\,0.01-0.09$\um and $\sim\,0.01-0.06$\um, respectively.  Although
the dispersion in the 3.3\um PAH EQW is large at a given 6.2\um PAH
EQW (especially at the high end of the 6.2\um PAH EQW), when
$\rm PAH(3.3\mu m)\,EQW > 0.04$\um, the EQWs of the 3.3\um and 6.2\um
PAH features can both be used to select starburst-dominated sources in
most cases.  However, LIRGs with $\rm PAH(3.3\mu m)\,EQW < 0.04$\um do
not show a decreasing $\rm PAH(6.2\mu m)\,EQW$ but instead display a
large dispersion in their 6.2\um PAH EQWs.  Based on their 3.3\um PAH
EQW alone, all of these galaxies would be classified as AGN-dominated,
while based on the 6.2\um feature at least $\sim 1/3$ ($7$ objects) of
them would still be classified as starburst galaxies (the green
squares in the top left gray region in
Figure~\ref{fig:PAH6_EQW_vs_PAH3_EQW}). Indeed, their AGN fractions
are $< 50$\%.  The galaxies with $\rm PAH(3.3\mu m)\, EQW < 0.04$\um
and $\rm PAH(6.2\mu m)\, EQW \geq 0.4$\um are \object{ESO~264-G036},
\object{ESO~319-G022}, \object{IC~0860}, \object{UGC~08739},
\object{NGC~6701}, \object{NGC~7591}, and \object{ESO~077-IG014}.  We
will investigate the nature of these sources in detail in
Section~\ref{subsec:discu:EQW}.

\begin{figure}
  \begin{center}
    \includegraphics[width=\textwidth/2]
    {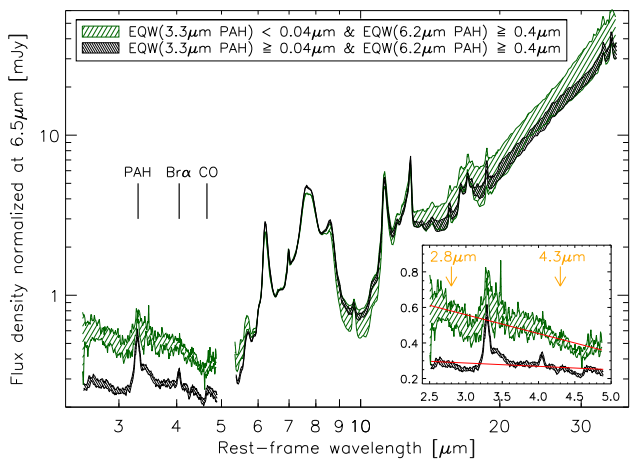}
    \caption{Median spectra of sources with
      $\rm PAH(6.2{\rm \mu m}) \, EQW \geq 0.4$\um, and
      $\rm PAH(3.3{\rm \mu m}) \, EQW < 0.04$\um (green; 7 objects) or
      3.3\,\um PAH EQW $\geq\,0.04$\um (black; 79 objects).  The
      ranges of the spectra are $1\sigma$ uncertainties of the
      estimated median values. The inset panel shows a zoom-in of the
      2.5-5\um region with the continuum fits in red.  The orange
      arrows indicate the wavelengths where the {\it AKARI} continuum
      color, $F_\nu(4.3{\rm \mu m})/F_\nu(2.8{\rm \mu m})$, is
      measured. }
    \label{fig:SED}
  \end{center}
\end{figure}

In Figure~\ref{fig:SED}, we show the median {\it AKARI} $+$ {\it
  Spitzer} spectra of galaxies with
$\rm PAH(6.2\mu m) \, EQW \geq 0.4$\um, but with
$\rm PAH(3.3\mu m) \, EQW < 0.04$\um (green) or
$\rm PAH(3.3\mu m) \, EQW \geq 0.04$\um (black).  We made a cut of
$\rm PAH(6.2\mu m) \, EQW$ at a more strict value of 0.4\um, instead
of 0.2\um, because we want to ensure a clean selection of starburst
dominated galaxies just by $\rm PAH(6.2\mu m) \, EQW$. We only use 6.2\um
PAH EQW to select starburst-dominated galaxies instead of the AGN
fraction because the scope here is to compare the spectra of galaxies
directly with their PAH EQWs.  The {\it AKARI} continuum of the median
spectrum with low $\rm PAH(3.3\mu m) \, EQW$ is bluer,
while the average {\it Spitzer} spectra are very similar.  When the
continua of the median {\it AKARI} spectra are fitted by a linear
function avoiding the $3.05$\um ($\rm H_2O$ ice) and $4.67$\um (CO ice
and gas) absorption features, the slopes are $-0.11$ and $-0.02$ for
sources with $\rm PAH(3.3\mu m) \, EQW < 0.04$\um and
$\rm PAH(3.3\mu m) \, EQW \geq 0.04$\um, respectively.
It is also evident from these median spectra that sources with high
3.3\um PAH EQW display a more prominent Br$\alpha$ emission.

\subsection{Ionized Gas in Starburst-Dominated LIRGs}\label{ss:ionized}

Because the intensity of hydrogen recombination lines is directly
related to the number of young, massive stars in star-forming regions,
they provide an accurate estimate of the current star formation rate.

Here we use the ratio of Br$\alpha _{4.05\,{\rm \mu
    m}}$/${\rm
  PAH}_{3.3\,{\rm \mu m}}$
to study the evolution of the ionization field in starburst-dominated
GOALS LIRGs.  In starburst-dominated galaxies, the Br$\alpha$ emission
arises from H{\sc ii} regions ionized by young stars, while the
continuum at 4\um is dominated by emission from old stars.  Thus, the
Br$\alpha$ EQW measures the light-averaged ratio of the emission from
young to old stars, 
and hence it can be used as an age indicator. Ages were estimated by
generating a single burst of star formation using the Starburst99
evolutionary synthesis models \citep[version
6.0.2;][]{Leit99,Vazq05,Leit10}. We used a Kroupa initial mass
function and the standard Geneva evolutionary tracks ($0.5-10\,$Myr
interval).  The model output of Br$\gamma$ EQW was converted to the
Br$\alpha$ EQW, assuming Case B for hydrogen recombination, an
effective temperature of $\sim\,10^4$~K, and a density~\footnote{ The
  median electron density of the entire GOALS sample is
  $\sim\,300\,{\rm cm^{-3}}$, estimated with the [S\,\textsc{iii}]
  mid-infrared emission line ratios \citep{Inam13}.} of
$\sim\,100\,{\rm cm^{-3}}$.  We also assume that there is just a
single ionizing population of stars producing both the line and
continuum emissions, which may not always be accurate, because in
reality an underlying older stellar population from past star-forming
episodes may also contribute to the near-infrared continuum.  Hence,
the ages derived from the Br$\alpha$ EQWs obtained with this simple
model are upper limits to the true age of the current starburst.

\begin{figure}
  \begin{center}
    \includegraphics[width=\textwidth/2]
   {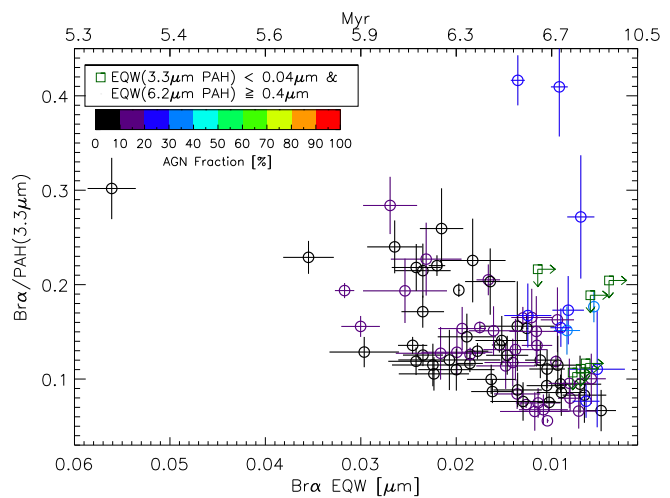}
   \caption{ The ratio of Br$\alpha$ to 3.3\um PAH emission as a
     function of decreasing Br$\alpha$ EQW (increasing starburst age,
     the upper x-axis).  The circles represent the sources with both
     3.3\um PAH and Br$\alpha$ detections color-coded by the AGN
     fraction. The upper limits are shown only for the objects with
     $\rm PAH(3.3{\rm \mu m}) \, EQW < 0.04$\um and
     $\rm PAH(6.2{\rm \mu m}) \, EQW \geq 0.4$\um.  The starburst ages
     are derived from the Br$\alpha$ EQW using the Starburst99 models
     \citep{Leit99,Vazq05,Leit10}. }
    \label{fig:Bra_PAH3_vs_BraEQW}
  \end{center}
\end{figure}

In Figure~\ref{fig:Bra_PAH3_vs_BraEQW}, we show the flux ratio of
Br$\alpha$ to ${\rm PAH}_{3.3\,{\rm \mu m}}$ as a function of
Br$\alpha$ EQW.  The range of Br$\alpha$ EQWs in the GOALS sample
corresponds to ages of $\sim\,5-10\,$Myr, with the majority of sources
being between 6 and 7\,Myr old.  These ages are slightly higher than
the ages of $1-4.5$~Myr of the GOALS LIRGs estimated from the
mid-infrared atomic fine-structure emission lines \citep{Inam13}, but
given that the Br$\alpha$ EQW ages are upper limits, they can be
considered consistent.  A clear correlation between
Br$\alpha$/${\rm PAH}_{3.3\,{\rm \mu m}}$ and Br$\alpha$ EQW is seen
in starburst-dominated sources (with an AGN fraction less than
50\%). The objects with $\rm PAH(3.3\mu m) \, EQW < 0.04$\um and
$\rm PAH(6.2\mu m) \, EQW \geq 0.4$\um (the green squares) are not
detected in Br$\alpha$.  The AGN-dominated sources (AGN fraction
$> 50$\%) do not follow the same trend: all of them have small
Br$\alpha$ EQWs and their Br$\alpha$/${\rm PAH}_{3.3\,{\rm \mu m}}$
ratios are between $\sim 0.1-0.25$.  Note that because an AGN can also
contribute to the hydrogen recombination line emission, the estimated
ages of the galaxies which have AGN present are probably not as
reliable.

For starburst-dominated sources, the decrease of the
Br$\alpha$/${\rm PAH}_{3.3\,{\rm \mu m}}$ ratio with decreasing
Br$\alpha$ EQW (increasing in starburst age) is consistent with the
aging of the ionizing stellar populations, and thus with the reduction
of Br$\alpha$ emission with stellar age. Our results are in agreement
with the findings of \cite{Diaz08,Diaz10a} who show that in resolved
star-forming regions of LIRGs, the ratio of the 11.3\,\um PAH (and
likely the 8.6\,\um PAH) to Pa$\alpha$ emission is correlated with the
age of the H{\sc ii} regions as traced by Pa$\alpha$ EQW. This trend
simply reflects the different time-scales probed by the ionized gas
and PAH emission, with the latter being also produced in more evolved
stellar populations (a few tens of Myr) and not ascribed only to
on-going star formation, because PAHs are excited by lower energy
non-ionizing photons.

\subsection{Contribution of PAH and Br$\alpha$ Emission to
  $L_{IR}$}\label{ss:ionized}

Both PAH and Br$\alpha$ emission trace ongoing star formation and are
often used to estimate SFRs.  Here, we examine the ratio of the
$3.3$\um PAH and Br$\alpha$ emission to the total infrared emission,
which can also be used to estimate the SFR in LIRGs, free of the
effects of extinction that can plague even the near-infrared emission.
The $3.3$\um PAH and Br$\alpha$ emission used in this section were
measured from the {\it AKARI} spectra that were extracted with the
aperture covering $\sim 95$\% of the total flux in the {\it AKARI} 2D
spectra (see Section~\ref{sec:data}).  This aperture was also employed
for the photometric measurements of the {\it Spitzer} MIPS imaging
data. The MIPS data were used as follows, in order to estimate the
infrared luminosity ($L_{IR}$) inside this aperture: First, a fraction
of the MIPS $24\,{\rm \mu m}$ flux density within this aperture and
the flux density of the entire galaxy were measured.  Then, this
fraction was utilized as a scaling factor to convert $L_{IR}$ of the
whole galaxy to $L_{IR}$ in the extraction aperture.  Hereafter, we
denote $L_{IR\,24}$ as the total infrared luminosity in this
extraction aperture.

\begin{figure}
  \begin{center}
    \includegraphics[width=\textwidth/2]
    {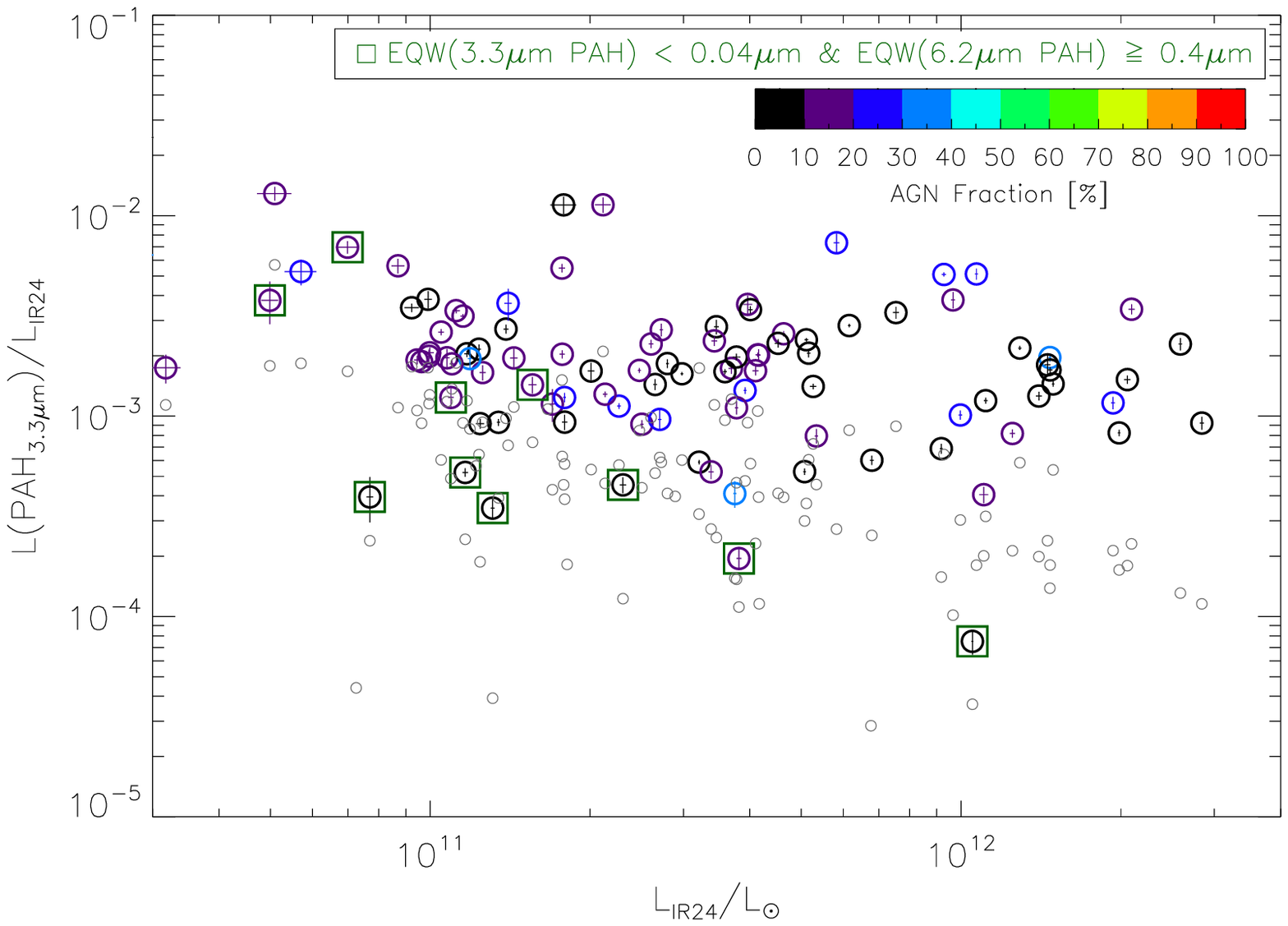}
    \includegraphics[width=\textwidth/2]
    {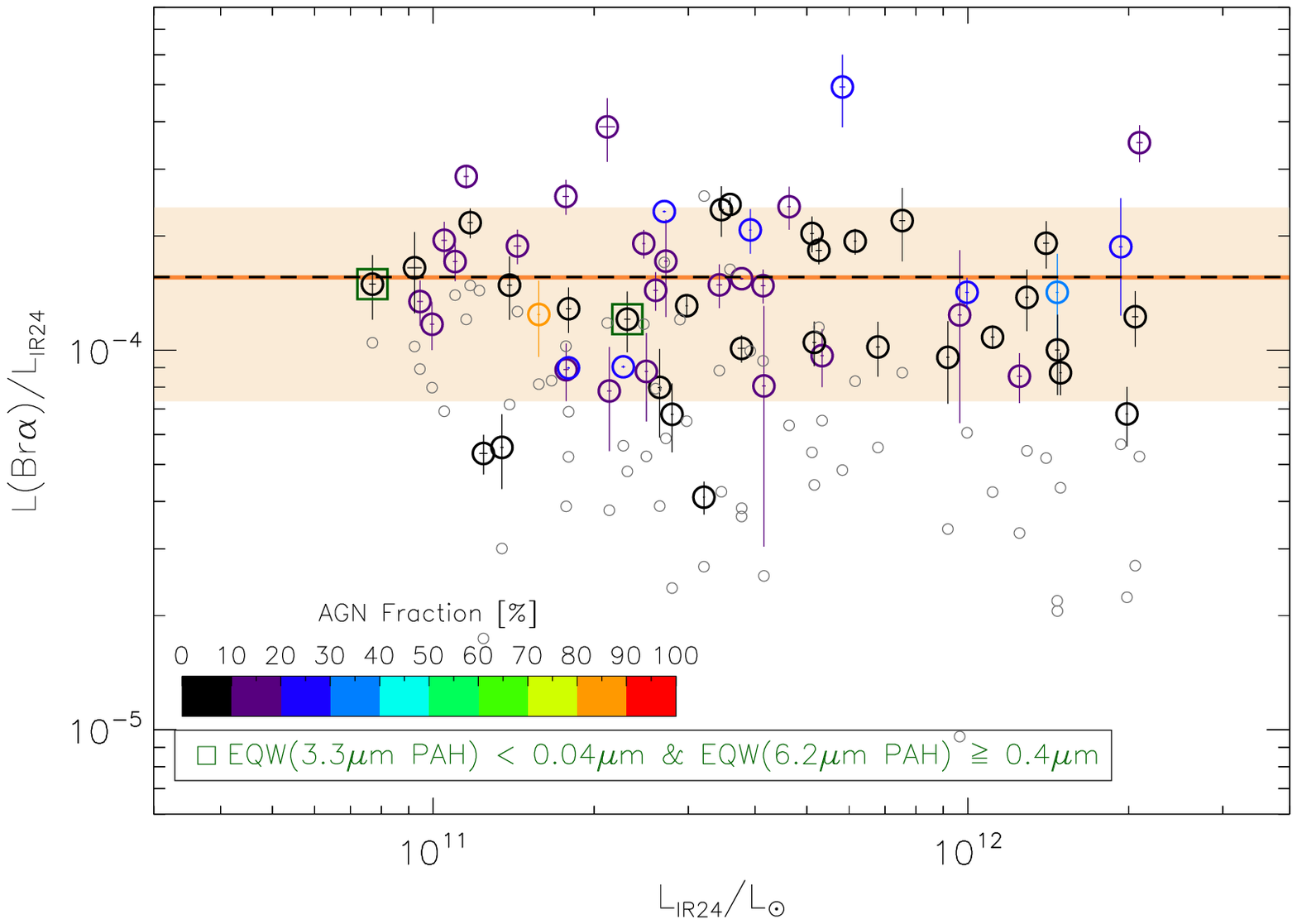}
    \caption{ [Top] The ratio of $L_{3.3\,{\rm \mu m} \, {\rm PAH}}$
      to $L_{IR\,24}$ as a function of $L_{IR\,24}$. The 3.3\,\um PAH
      emission is measured from the {\it AKARI} spectra extracted with
      the aperture that covers $\sim\,95$\% of the total flux in the
      {\it AKARI} 2D spectra.  $L_{IR\,24}$ is the total infrared
      luminosity, scaled by the flux ratio of the {\it Spitzer}/MIPS
      $24$\um emission within the {\it AKARI} aperture to the total
      $24$\um flux (see Section~\ref{ss:ionized} for more details).
      The sources with the detected emission feature are shown by the
      circles.  The small gray circles denote the values without the
      dust extinction correction.  The large circles are color-coded
      by the AGN fraction.  The green squares represent the sources
      with $\rm PAH(3.3{\rm \mu m}) EQW < 0.04$\um and
      $\rm PAH(6.2{\rm \mu m}) EQW \geq 0.4$\um.  [Bottom] The ratio
      of measured (gray) and corrected (colored) Br$\alpha$ to the
      infrared luminosity as a function of $L_{IR\,24}$.  The
      horizontal orange solid line and shaded area indicate the mean
      and its standard deviation of the extinction corrected
      $L_{\rm Br\alpha}/L_{IR\,24}$, respectively.  The horizontal
      black dashed line indicates
      $L_{\rm Br\alpha}/L_{IR} = 1.56 \times 10^{-4}$ when the SFRs
      derived from $L_{\rm Br\alpha}$ and $L_{IR\,24}$ are the same.}
    \label{fig:PAH3_LIR_vs_LIR_phot} 
  \end{center}
\end{figure}

In the top panel of Figure~\ref{fig:PAH3_LIR_vs_LIR_phot}, we present
the ratio of the $3.3{\rm \mu m}$ PAH luminosity to $L_{IR\,24}$ as a
function of $L_{IR\,24}$.  We show both the dust extinction
uncorrected (small gray circles) and corrected (colored circles) PAH
luminosities.  The dust extinction is estimated based on the optical
depth of the 9.7\um silicate feature in the {\it Spitzer}/IRS spectra
using Eq.(4) of \cite{Smit07}. The measurements of
$\tau(9.7\,{\rm \mu m})$ are taken from \cite{Stie14}.  The
$\tau(9.7\,{\rm \mu m})$ range of our sample is $0.46-8.38$ with the
mean and median of 1.92 and 1.57, respectively.  There is a trend of
decreasing $L_{3.3{\rm \mu m\,PAH}}/L_{IR\,24}$ (a power-law function
with an exponent of $-0.55$), indicating that SFR based on uncorrected
$L_{3.3{\rm \mu m\,PAH}}$ is likely to be underestimated for more
infrared luminous galaxies.  This trend has already been seen in the
past for local LIRGs, which exhibit a deficit in
$L_{3.3{\rm \mu m\,PAH}}/L_{IR}$ (as well as in
$L_{6.2{\rm \mu m\,PAH}}/L_{IR}$) compared with less infrared luminous
galaxies \citep[e.g.,][]{Kim12,YamaR13,Mura17,Stie14}.  Although the
extinction correction for PAH based SFR calibrations is not common in
the literature \citep[e.g.,][]{Calz07, Kenn09, Ship16}, when the dust
extinction correction is applied, this trend becomes weaker, but it is
still evident as a power-law function with an exponent of $-0.24$.
All extreme outliers have $\rm PAH(3.3\mu m) \, EQW < 0.04$\um.

As shown in the bottom panel of Figure~\ref{fig:PAH3_LIR_vs_LIR_phot},
the mean and its standard deviation of the measured extinction
corrected $L_{\rm Br\alpha}/L_{IR\,24}$ are
$(1.55 \pm 0.82) \times 10^{-4}$, indicating that the
SFR($L_{IR\,24}$) and the extinction corrected SFR(Br$\alpha$) are
consistent \citep{Kenn98b}.  For the most infrared luminous galaxies,
it is particularly important to perform the dust extinction correction
based on the 9.7\um absorption feature to bring the two measures in
line across the sample \citep[see also][]{daCu10}.


\section{Discussion}\label{sec:discussion}

Based on our measurements of the combined {\it AKARI} and {\it
  Spitzer} spectra of 130 local LIRGs\,\footnote{There are 15 sources
  which are not included in the direct comparisons between their {\it
    AKARI} and {\it Spitzer} data because of mismatched pointings or
  their extended morphologies (Section~\ref{sec:results}).}, we
examine the nature and properties of these galaxies from the GOALS
sample. In this section, we propose an improved starburst/AGN
diagnostic and explore ionization state and size distribution of the
dust.

\subsection{Nature of Low 3.3\,\um PAH EQW Starburst
  LIRGs}\label{subsec:discu:EQW}

An empirical cut at $\rm PAH(3.3\mu m) \, EQW < 0.04$\um has often
been used to identify AGN-dominated sources
\citep[e.g.,][]{Moor86,Iman10,YamaR13}. However, as we have shown in
Section~\ref{ss:paheqws}, out of 19 (or 24 including upper limits)
sources with 3.3\,\um PAH EQW in this range, $\gtrsim 1/3$\ (7) have
$\rm PAH(6.2\mu m) \, EQW \geq 0.4$\um, which implies that they are
starburst-dominated galaxies.  Based on the mid-infrared AGN
diagnostics discussed in Sec~\ref{ss:paheqws}, none of these sources
show any indication of AGN dominance even with the individual
mid-infrared AGN indicators. Furthermore, close inspection of
available archival optical spectra also indicates that they do not
harbour an AGN. There is clearly a class of sources where the 3.3\um
PAH EQW alone can lead to erroneous identification of AGN, since it is
very unlikely that the dust emission from an AGN dominates the
$2-5$\um wavelength range but does not contribute significantly to the
mid-infrared ($T\sim\,500\,$K), given their intrinsically red,
power-law continuum emission \citep{Weed05,Armu07}.  The following
three factors may contribute to having a LIRG with a low EQW for
3.3\um PAH ($< 0.04$\um) while the corresponding 6.2\um PAH EQW
remains $> 0.4$\um: (1) an absence of small PAH molecules, (2) a
larger fraction of ionized PAH molecules, or (3) an excess of stellar
emission at $\lambda \lesssim 5 $\um.  We discuss these hypotheses
below.

Laboratory and numerical studies have shown that the relative
strengths of the different PAH bands are expected to vary with the
size and the ionization state of PAH molecules
\citep{Drai01,Li01,Drai07a,Tiel08}. This is caused by the different
absorption cross section per C atom of neutral and ionized PAH (see
Figure~2 of \cite{Li01} and Figure~3 of \cite{Drai07a}). Ionization
enhances the C-C stretching modes (6.2 and $7.7$\um) and the C-H
in-plane bending mode ($8.6$\um), but weakens the C-H stretching mode
(3.3\um) and the C-H out-of-plane bending mode ($11.3$\um).  In
addition, PAH features for smaller grain sizes have emission peaks at
shorter wavelengths. In other words, at a certain ionization state,
when large dust grains are the predominant population, the 8.6\um PAH
feature shows stronger emission than the 6.2\um PAH.  Similarly, a
harder ionization field produces a larger fraction of ionized PAH,
such that the 3.3\um PAH feature (and $11.3$\um PAH) is weaker,
whereas the 6.2\um PAH feature (as well as the $7.7$\um and $17$\um
PAH features) is stronger. Thus, the PAH emission ratios such as
${\rm PAH}_{3.3{\rm \mu m}}/{\rm PAH}_{6.2{\rm \mu m}}$,
${\rm PAH}_{11.3{\rm \mu m}}/{\rm PAH}_{6.2{\rm \mu m}}$, and other
combinations of ionized and neutral PAH emission can be used as
indicators of the PAH ionization state.  For example, \cite{Jobl96}
have found that the
${\rm PAH}_{8.6{\rm \mu m}}/{\rm PAH}_{11.3{\rm \mu m}}$ ratio depends
on the relative populations of ionized and neutral PAH, which is
supported by laboratory experiments \citep{Hudg95}.

We use the flux ratio-ratio diagram of
${\rm PAH}_{11.3{\rm \mu m}}/{\rm PAH}_{7.7{\rm \mu m}}$ and
${\rm PAH}_{6.2{\rm \mu m}}/{\rm PAH}_{7.7{\rm \mu m}}$ as in Figure~2
of \cite{Stie14} to investigate where the sources with low 3.3\um and
high 6.2\um PAH EQWs lie compared with the entire sample.  The PAH
fluxes from the {\it Spitzer} data are also taken from \cite{Stie14}.
The sources with the low 3.3\um and high 6.2\um PAH EQWs (i.e. the
squares in Figure~\ref{fig:PAH6_EQW_vs_PAH3_EQW}) are distributed over
a wide range of the grain size and ionization state without any
obviously biased distributions compared to the other sources. In
addition, these sources do not show particularly weak 3.3\um PAH
emission relative to 6.2\um PAH emission.  A direct flux comparison of
the 3.3\um and 6.2\um PAH features of these sources is consistent with
the distribution of the rest of our sample, except for
\object{IC~0860} whose $F_{\rm 6.2\,PAH}/F_{\rm 3.3\,PAH}$ ratio is
about a factor of five higher than the average of the entire sample.
Therefore, the fact that they (possibly except \object{IC~0860}) have
low 3.3\um and high 6.2\um PAH EQWs is related to the continuum
properties, not the PAH properties.

\begin{figure}
  \begin{center}
    \includegraphics[width=\textwidth/2]
    {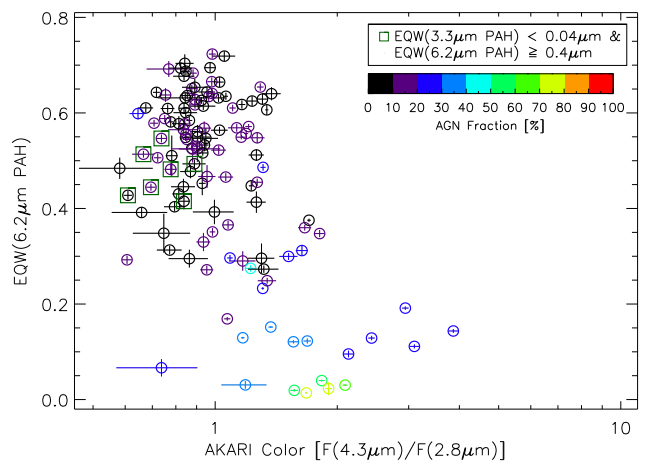}
    \caption[]{The PAH(6.2\um) EQW plotted against the {\it AKARI}
      $2.5-5$\um continuum slope as measured by the
      $F_\nu(4.3{\rm \mu m})/F_\nu(2.8{\rm \mu m})$ ratio (see also
      Figure~\ref{fig:SED}).  The symbols are the same as in
    Figure~\ref{fig:PAH6_EQW_vs_PAH3_EQW}.  }
    \label{fig:PAH_EQW_vs_NIR_Color}
  \end{center}
\end{figure}

Next, we examine the $6.2$\um PAH EQW versus the $2-5$\um continuum in
Figure~\ref{fig:PAH_EQW_vs_NIR_Color}.  With one exception, all LIRGs
with AKARI flux density ratios
$F_\nu(4.3{\rm \mu m})/F_\nu(2.8{\rm \mu m}) \lesssim 1$ display
$\rm PAH(6.2\mu m) \, EQW > 0.2$\um.  When only LIRGs with
$\rm PAH(6.2\mu m) \, EQW < 0.4$\um are considered, sources with
lower PAH EQWs tend to show progressively redder continuum
slopes. More strikingly, galaxies for which there is inconsistency
between the 3.3\um and 6.2\um PAH EQW starburst/AGN classifications
(i.e., those with $\rm PAH(6.2\mu m) \, EQW \geq 0.4$\um and
$\rm PAH(3.3\mu m) \, EQW < 0.04$\um, the circles with the green
square) occupy the same region as those for which both diagnostics
indicate that they are dominated by star formation. That is, they have
$2-5$\um continuum slopes as blue as or bluer
($F_\nu(4.3{\rm \mu m})/F_\nu(2.8{\rm \mu m}) \lesssim 0.9$) than most
of the pure star-forming LIRGs. The bluer slope can also be identified
in Figure~\ref{fig:SED}.

\begin{figure}
  \begin{center}
    \includegraphics[width=\textwidth/2]
   {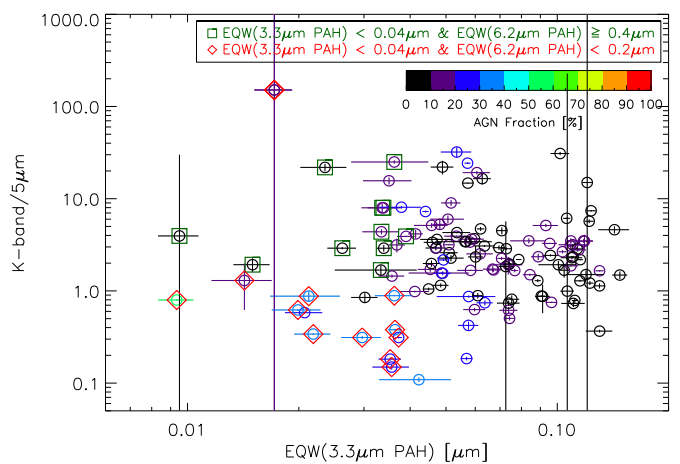}
   \caption{ Ratio of {\it 2MASS} $K$-band to $5$\um monochromatic
     continuum emission as a function of 3.3\,\um PAH EQW.  The EQWs
     are derived from the {\it AKARI} spectra extracted with an
     aperture covering $\sim\,95\%$ of the total {\it AKARI} flux. The
     $K$-band emission is measured in the imaging data using the same
     aperture size.  The symbols are the same as in
     Figure~\ref{fig:PAH6_EQW_vs_PAH3_EQW}, in addition to the red
     diamonds, which indicate $\rm PAH(3.3{\rm \mu m}) EQW < 0.04$\um
     and $\rm PAH(6.2{\rm \mu m}) EQW < 0.4$\um.}
    \label{fig:K_5_vs_EQW3}
  \end{center}
\end{figure}

Therefore, the $2-5$\um continua in some of the LIRGs with the lowest
3.3\um PAH EQWs are not necessarily dominated by hot dust emitted by
the optically thick dusty torus of a putative central AGN.  To explore
this further, in Figure~\ref{fig:K_5_vs_EQW3}, we plot the ratio of
$K$-band, a first order tracer of stellar mass of the entire galaxy,
to 5\um monochromatic emission, as a function of the 3.3\um PAH EQW to
probe the dominant source of the continuum emission. Galaxies with
$\rm PAH(3.3\mu m) \, EQW < 0.04$\um roughly split into two different
populations depending on their $K$-band/5\um ratios.  Among these
galaxies, interestingly, when their 6.2\um PAH EQWs are high
($\geq 0.4$\um, green squares), they also show higher $K$-band/5\um
ratios ($\gtrsim 1.0$). On the other hand, the galaxies with
$\rm PAH(6.2\mu m) \, EQW < 0.2$\um (red diamonds) have $K$-band/5\um
$\lesssim 1.0$. 
The higher $K$-band/5\um ratio for the galaxies with the low 3.3\um
PAH and high 6.2\um PAH EQWs indicates a larger stellar contribution
at $K$-band which is likely produced by the stellar continuum of more
massive, evolved stars rather than hot dust.

Dust extinction cannot cause the discrepancy between the PAH
EQW(3.3\um) and EQW(6.2\um) diagnostics, because it would also
suppress the 3.3\um PAH flux with respect to the 6.2 PAH flux.
However, if the $\sim 3$\um continuum is dominated by stellar emission
instead of hot dust and the $\sim 3$\um continuum emission is more
extended than the 3.3\um PAH flux, this may decrease the PAH
EQW(3.3\um) with respect to the PAH EQW(6.2\um).  Because the IRS slit
is narrow, we inspect these galaxies to check whether they are more
distant, such that more emission from galaxy disks is covered by the
slit. We find that their luminosity distances are around the median
distance of the entire sample, except \object{ESO~077-IG014} which is
at 186~Mpc. Therefore, while stellar emission could be the source of
the decreased 3.3\um PAH EQW, a greater distance is not responsible
for the excess disk emission in these sources.

These analyses suggest that stellar emission plays an important role
in some LIRGs.  While these galaxies are actively star-forming, there
is very little hot dust contributing to their observed 3\um emission.
This hot dust could be highly compact and obscured, or these galaxies
may simply have less hot dust because of a more distributed or aging
nuclear burst.

\subsection{Further Constraints on the {\it AKARI} Starburst/AGN
  Diagnostics}\label{sec:new_diagnositcs}

\begin{figure*}
  \begin{center}
    \includegraphics[width=0.8\textwidth]
    {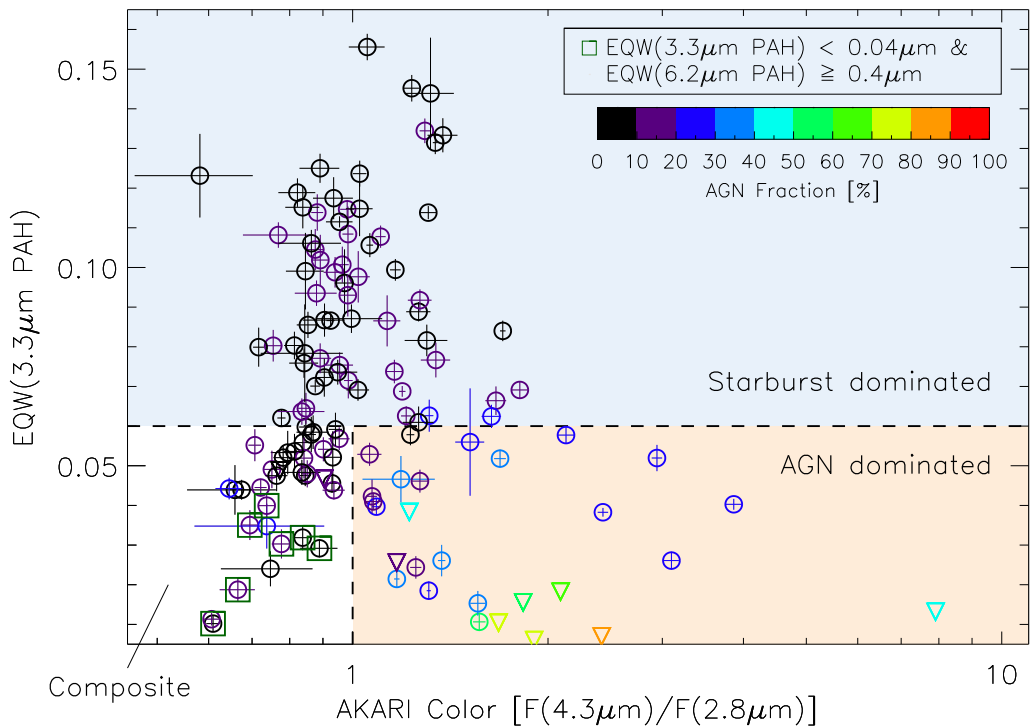}
    \caption{ Revised starburst/AGN diagnostic diagram using 3.3\um
      PAH EQW and $F_\nu(4.3{\rm \mu m})/F_\nu(2.8{\rm \mu m})$.  The
      symbols are the same as in
      Figure~\ref{fig:PAH6_EQW_vs_PAH3_EQW}.  The blue and beige
      areas, respectively, represent the regions that we propose to
      identify LIRGs for which starburst and AGN dominate the
      continuum emission in the infrared (see also Eq.~\ref{eq:SB} and
      ~\ref{eq:AGN}). The white area denotes composite sources
      (Eq.~\ref{eq:comp}).  }
    \label{fig:AKARI_diagnostics}
  \end{center}
\end{figure*}

Several studies have proposed using the 3.3\um PAH EQW or the $2-5$\um
continuum slope ($\Gamma$; $F_\nu\propto\lambda^\Gamma$) from {\it
  AKARI} as diagnostics for the identification of obscured AGN in
LIRGs \citep{Iman10,Lee12}.  The parameter space within which galaxies
harboring an AGN are found, is normally assumed to have
$\rm PAH(3.3\mu m) \, EQW < 0.04$\um or
$\Gamma_{(F_\lambda\propto\lambda^\Gamma)} > -1$ \citep[equivalent to
$\Gamma_{(F_\nu\propto\lambda^\Gamma)} > 1$,][]{Risa10,Sani08,Iman10}.
However, these studies are limited to the information available at
$2-5$\um, which is where the hottest dust emission from the obscured
AGN ($T > 1000$K) and direct photospheric starlight are in competition
with each other.

The mid-infrared is a much cleaner wavelength regime to search for
buried AGN \citep{Laur00, Stur02, Armu07, Desa07, Petr11,
  Alon12}. This is where warm ($T \sim 300-500$K) dust from the torus
can contribute significantly to the continuum emission \citep{Desa07,
  Spoo07, Petr11,Diaz11}.  

Based on the additional information provided by the {\it Spitzer}/IRS
data, we propose a revision of the commonly used starburst/AGN
diagnostic in the $2-5$\um range. This involves the 3.3\um PAH EQW and
the continuum slope, $\Gamma$, which we trace using the flux density
ratio, $F_\nu(4.3{\rm \mu m})/F_\nu(2.8{\rm \mu m})$.  We prefer the
flux ratio because $\Gamma$ requires fitting the continuum slope, and
a wide variation in the slope within the $2-5$\um range
(Figure~\ref{fig:akari_spec}) can easily cause a unreasonable fits
(and corresponding $\Gamma$ values).

We propose revised boundaries between starburst- and AGN-dominated
galaxies as follows:

\begin{itemize}
  \item Starburst-dominated sources:
    \begin{equation}
      {\rm PAH(3.3{\rm \mu m})\,EQW} \geq 0.06 {\rm \mu m}
      \label{eq:SB}
    \end{equation}
  \item AGN-dominated sources:
    \begin{equation}
      \left\{
        \begin{array}{l} 
          {\rm PAH(3.3{\rm \mu m}) \, EQW} < 0.06 {\rm \mu m} \\ 
          F_\nu(4.3{\rm \mu m})/F_\nu(2.8{\rm \mu m}) \geq 1.0
        \end{array}
      \right.
      \label{eq:AGN}
    \end{equation}
\end{itemize}

We display this new diagnostic for the GOALS AKARI sample in
Figure~\ref{fig:AKARI_diagnostics}.  In most cases, the 3.3\um PAH EQW
can be used to select starburst-dominated galaxies.  Note that while
many other studies have drawn similar limits on these measurements to
derive near-infrared diagnostics for AGN searches, they use these two
limits (the PAH EQW and color) independently. We require instead that
both quantities be taken into account. Otherwise as much as $1/3$ of
our LIRG sample with $\rm PAH(3.3\mu m) \, EQW < 0.04$\um could be
misclassified as AGN when they are more likely to be
starburst-dominated galaxies.

Sources in which both star formation and AGN may contribute
significantly to the near- and mid-infrared emission are:

\begin{itemize}
  \item Composite sources:
    \begin{equation}
      \left\{
        \begin{array}{l} 
          {\rm PAH(3.3\,{\rm \mu m})~EQW}\,<\,0.06\,{\rm \mu m} \\ 
          F_\nu\,(4.3\,{\rm \mu m})/F_\nu\,(2.8\,{\rm \mu m})\,<\,1.0
        \end{array}
      \right.
      \label{eq:comp}
    \end{equation}
\end{itemize}

We cannot completely rule out the possibility that LIRGs which meet
the conditions of Eq.(\ref{eq:comp}) are AGN. Nevertheless, as
shown in Section~\ref{subsec:discu:EQW}, at least the near and
mid-infrared emission of these LIRGs is likely dominated by star
formation processes, which implies that AGN probably do not contribute
significantly to the bolometric luminosity of these galaxies. In fact,
their AGN fractions are $\lesssim 50\%$.

\subsection{Applications to {\it JWST} Observations}\label{subsec:discu:JWST}

\begin{figure*}
  \begin{center}
    \includegraphics[width=0.8\textwidth,trim=0 0 20 10,clip]
    {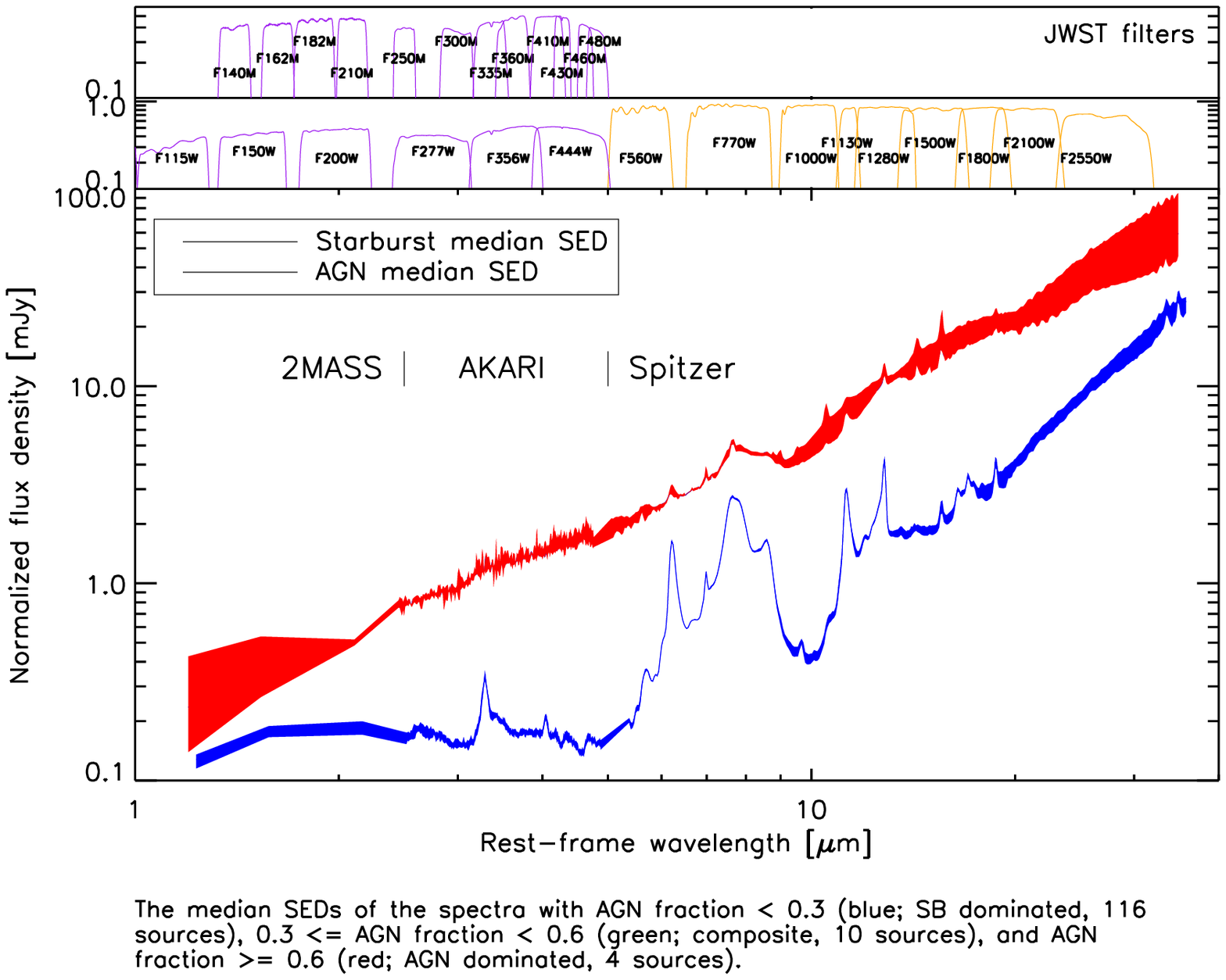}
    \caption{ [Top] The {\it JWST} filter response curves. The purple
      and orange curves denote the NIRCam and MIRI filters,
      respectively. [Bottom] The median SEDs of starburst- (blue) and
      AGN-dominated (red) LIRGs.  The starburst- and AGN-dominated
      galaxies are selected to have the AGN fractions $< 30\%$ and
      $\geq 60\%$, respectively. The SEDs are shown in the
      rest-frame. The median fluxes are manually scaled to avoid any
      overlap between the SEDs. The $1.2-2.2$\um, $2.5-5$\um, and
      $5-38$\um data are from 2MASS, {\it AKARI}, and {\it Spitzer},
      respectively.  The redshifted SEDs ($z=0.3-5.0$) are shown in
      Figure~\ref{fig:JWST_SED_z}.  }
    \label{fig:JWST_SED}
  \end{center}
\end{figure*}

With the $2.5-38$\um spectra of a large sample of local LIRGs, we are
able to investigate the effectiveness of using the existing {\it
  JWST}/NIRCam and {\it JWST}/MIRI filters to select starburst and
AGN.  In order to fully cover the {\it JWST} wavelength range, we have
also measured the $J$, $H$, and $K_S$-band fluxes using the imaging
data from the Two Micron All Sky Survey \citep[2MASS,][]{Skru06}. The
centroid of the photometry aperture is the same as the {\it Spitzer}
pointing, which is also the spectral extraction position of the {\it
  AKARI} data.  The aperture size is $10.5\arcsec$ to match the {\it
  AKARI} and {\it Spitzer} spectral extractions (see
Section~\ref{sec:data}).  The median SEDs of starburst and
AGN-dominated LIRGs from our sample are shown in
Figure~\ref{fig:JWST_SED}, along with the {\it JWST} filter response
curves for reference.  To create these median spectra, we combine all
LIRGs with AGN fractions less than 30\% as starburst-dominated, and
all those with AGN fractions more than 60\% as AGN dominated.  In the
following subsections, the $1.2-38$\um SEDs of individual galaxies and
the median SEDs of the starburst and AGN-dominated galaxies shown in
Figure~\ref{fig:JWST_SED} are employed to explore AGN color selections
and $3.3$\um PAH spectro-photometry using the {\it JWST} photometric
filters.

\subsubsection{AGN Color Selections}

\begin{figure}
  \begin{center}
    \includegraphics[width=\textwidth/2,trim=15 -5 90 -290,clip]
    {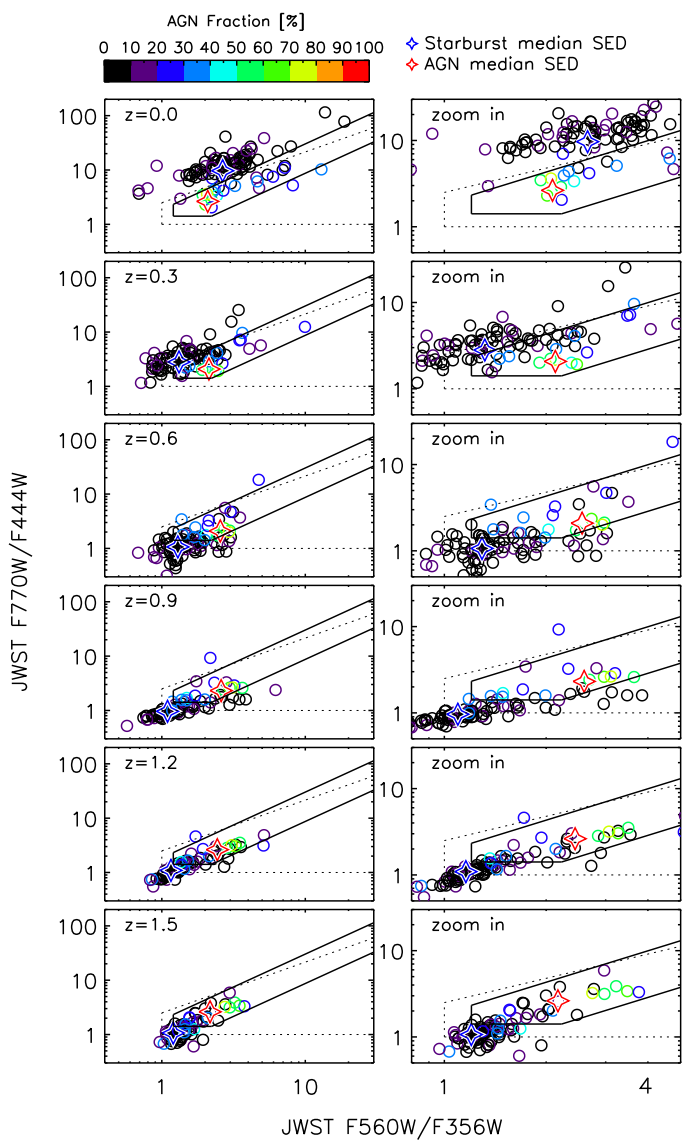}
    \caption{ The AGN infrared color selection with the {\it JWST}
      filters adopted from \cite{Lacy04} (the dotted line) and
      \cite{Donl12} (the solid line). The circles indicate individual
      LIRGs with the color coding of the AGN fraction shown on the top
      of the figure.  The blue (starburst) and red (AGN) stars
      indicate the median SEDs from Figure~\ref{fig:JWST_SED}. The
      panels show the changes of the galaxy colors with the redshift
      varying from the top ($z=0.0$) to the bottom ($z=1.5$) in steps
      of $\Delta z = 0.3$.  The left column shows the full range of
      the galaxy color distribution, while the right column shows a
      magnified portion of the left plot. The plots are shown only up
      to $z = 1.5$ due to the wavelength limit (2MASS $J$-band) of our
      SEDs being shifted beyond the coverage of F356W.}
    \label{fig:JWST_Lacy_evol}
  \end{center}
\end{figure}

In Figure~\ref{fig:JWST_Lacy_evol}, we show a color-color diagram of
the {\it JWST} filters F770W/F444W vs. F560W/F356W. These broadband
filters are selected to be closest to the color combinations of the
{\it Spitzer}/IRAC filters for the AGN selection in \cite{Lacy04} and
\cite{Donl12}.  We adapt the same IRAC color boundaries for our {\it
  JWST} color-color diagram.  \cite{Donl12} have explored the color
evolution of galaxies with various AGN contributions using the
redshifted SED templates of \cite{Poll08} and \cite{Dale02} in the
IRAC color-color diagram.  Here we artificially redshift our combined
2MASS, {\it AKARI}, and {\it Spitzer} SEDs from $z=0$ to $z=1.5$ in
steps of $\Delta z = 0.3$ to explore their color evolution in the {\it
  JWST} color-color diagram.

As shown in Figure~\ref{fig:JWST_Lacy_evol}, AGN-dominated galaxies
are mostly found within the AGN selection boundaries of both
\cite{Lacy04} and \cite{Donl12} up to $z=1.5$, while
starburst-dominated galaxies become bluer in both of the colors from
$z=0$ to $0.3$.  At higher redshifts, starburst-dominated galaxies
enter and remain within the AGN boundaries of \cite{Lacy04} and cannot
be separated from the AGN-dominated galaxies. On the other hand, the
boundaries of \cite{Donl12} give a cleaner selection as expected.
This trend agrees well with the results of \cite{Donl12} that pure
starburst galaxies at $0.5 < z < 1.5$ become bluer and lie on the
bottom left side just outside of the AGN selection boundaries. For
this color-color selection, we can only perform the test up to $z=1.5$
because of the wavelength coverage of our SEDs, whose blue-end (2MASS
{\it J}-band) shifts out from the coverage of F356W.

In addition to adapting the IRAC color selections, we also explore
other possible color selection criteria based entirely on the {\it
  JWST} filters.  The best approach would be to use the NIRCam
medium-band filters to isolate the 3.3\um PAH EQW using the
F335M/F300M flux ratio and detect hot dust emission using F430M/F250M
for galaxies at $z \sim 0.4$.  However, the red wavelength limit of
the medium-band filters at F480M prevents us from applying this method
for galaxies at higher redshifts ($z \geq 0.4$).  We have tried the
same experiment using the broadband filters (e.g., F356W/F277W
vs. F444W/F200W for $z=0$) in order to apply the same selection method
to higher redshifts. However, this does not provide a clean separation
between AGN- and starburst-dominated sources, due to the intrinsically
low PAH EQW.

\begin{figure}
  \begin{center}
    \includegraphics[width=\textwidth/2,trim=50 20 20 -330,clip]
    {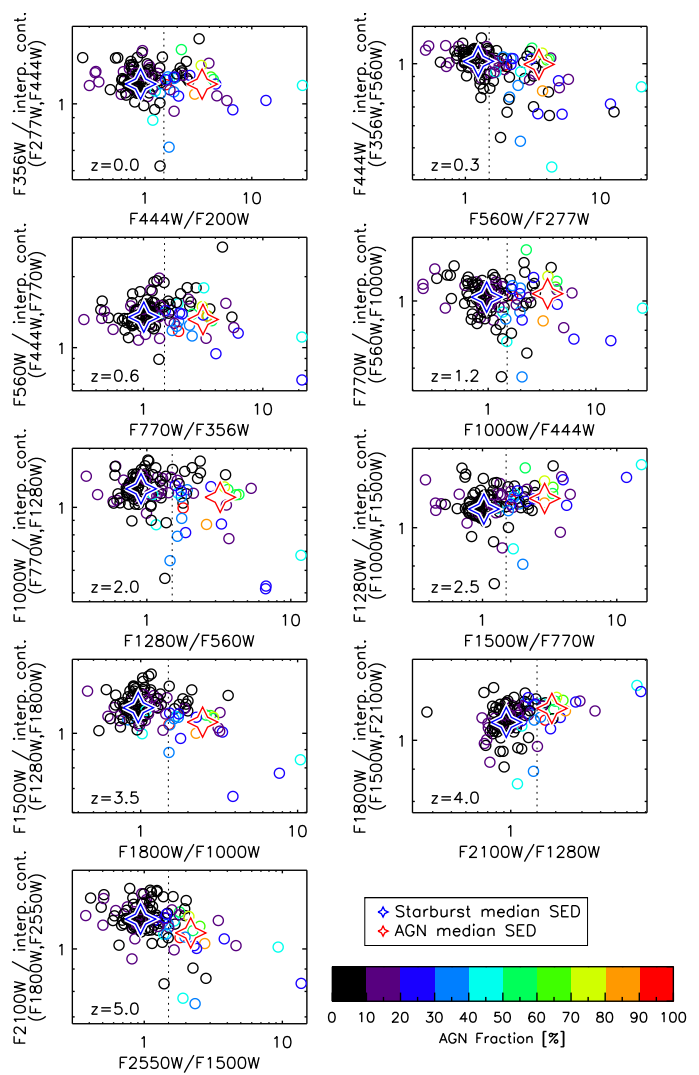}
    \caption{ AGN/starburst diagnostic diagrams at each of the
      redshifts indicated at the bottom left of each panel.  The
      y-axis represents 3.3\um PAH EQW by taking the flux ratio of the
      filter covering the 3.3\um PAH feature to the estimated local
      continuum emission via a linear interpolation between the two
      bracketing filters. The x-axis is a measure of hot dust
      emission. The symbols are the same as in
      Figure~\ref{fig:JWST_Lacy_evol}. The vertical dotted line at the
      colors of $1.5$ in each panel indicates an empirical cut for
      selecting starburst-dominated galaxies.  }
    \label{fig:JWST_AGN_z}
  \end{center}
\end{figure}

Instead of using a simple color to represent the PAH EQW, we assume
that the local continuum of the $3.3$\um PAH feature is the linear
interpolation of the two filters that bracket the filter which covers
the $3.3$\um PAH feature to give an estimate of 3.3\um PAH EQW
(e.g., at $z=0$, the PAH is in F336W, and F277W and F444W are used to
compute its local continuum).  In Figure~\ref{fig:JWST_AGN_z}, the
ratio of the flux in the filter band of the PAH feature to the
estimated local continuum is used to approximate the 3.3\um PAH EQW
(y-axis). This represents a reproduction of
Figure~\ref{fig:AKARI_diagnostics} using the {\it JWST} filters.  The
flux ratio along the x-axis indicates the continuum slope which traces
the strength of hot dust emission. Due to relatively low 3.3\um
emission compared to the broadband filters, it can be challenging to
accurately estimate actual EQW values, but the hot dust indicator
still works well to identify dust embedded AGN.

For the galaxies with the rest-frame $2-5$\um color (x-axis) of
$< 1.5$, all of them are starburst dominated (with AGN fraction less
than 50\%). All of the AGN-dominated sources have the color of
$> 1.5$, as well as do some of the starbursts.  Among all of the
starburst sources, about $20-30\%$ (except at $z=0.3$, which is 45\%)
have the colors of $\geq 1.5$.  While selections of AGN-dominated
sources with the colors at $\geq 1.5$ at each redshift include some
contaminants, the color limit of $\lesssim 1.5$ can act as a
reasonable empirical limit for selecting starburst-dominated galaxies
up to $z=5$.

\subsubsection{PAH Flux Inferred from {\it JWST} Broadband Photometry}\label{ss:JWST_excess}

Here we examine whether we can use the {\it JWST} broadband filters to
estimate the $3.3$\um PAH flux by identifying a color excess.  Using
our sample, we estimate the flux of the 3.3\um PAH emission assuming
$L_{3.3{\rm \mu m \, PAH}}/L_{IR} \sim 0.1\%$
(Figure~\ref{fig:PAH3_LIR_vs_LIR_phot} top).  For a system with
$L_{IR} = 10^{12} \, L_\odot$, the expected 3.3\um PAH fluxes are
$3.2 \times 10^{-15} \, {\rm W \, m^{-2}}$ at $z \sim 0$,
$4.6 \times 10^{-19} \, {\rm W \, m^{-2}}$ at $z \sim 1.2$, and
$1.4 \times 10^{-20} \, {\rm W \, m^{-2}}$ at $z \sim 5$.  Based on
the unresolved line sensitivity of {\it JWST}/MIRI spectroscopy
\citep{Glas15}, this suggests that the feature would be detectable out
to $z \sim 4.5$ at $10\sigma$ within 10000 sec of on-source exposure
time.

Following the methodology described in the previous section, we assume
a linear interpolation of two neighboring filters to represent the
local continuum under the 3.3\um PAH feature.  An advantage of using
$3.3$\um PAH is that the local continuum is often flatter than the
other PAH features and rarely has strong absorption or emission
features around $3.3$\um to interfere with measuring the excess in a
broadband filter.  In addition, although the $3.3$\um PAH EQW is on
average about a factor of 10 smaller than $6.2$\um PAH EQW, it is the
only PAH feature that can be observed with {\it JWST} at $z > 3.5$.

\begin{figure}
  \begin{center}
    \includegraphics[width=\textwidth/2,trim=14 30 20 -330,clip]
    {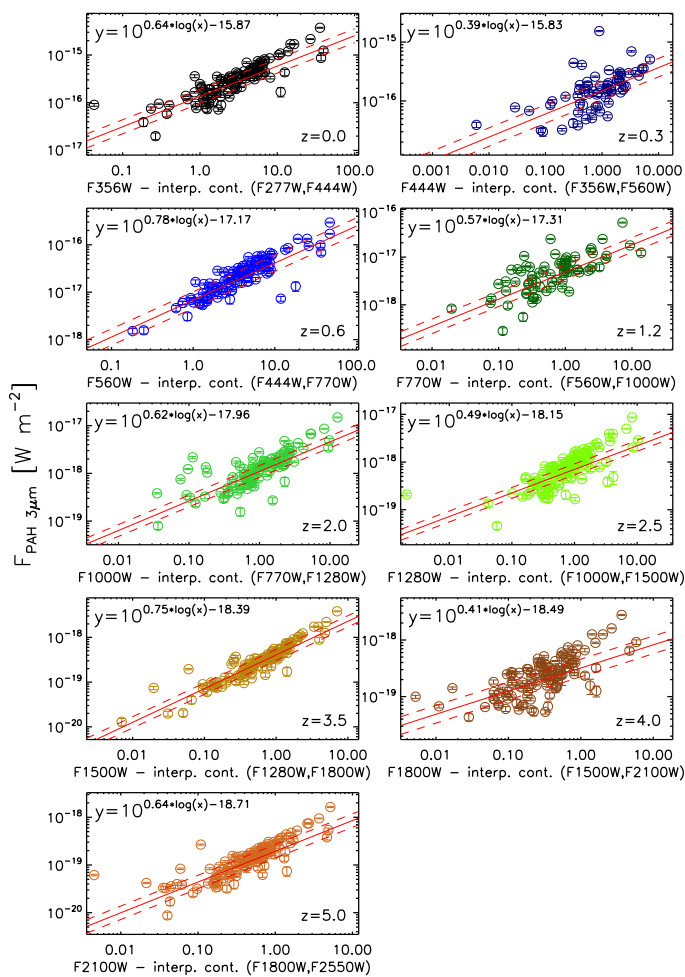}
    \caption{ The 3.3\um PAH fluxes obtained in the {\it AKARI}
      spectra compared with an estimate of the 3.3\um PAH fluxes using
      three {\it JWST} broadband filters from $z=0$ to $5.0$. Note
      that at $z=0$, the 3.3\um PAH fluxes are the actual measurements
      from the {\it AKARI} spectra, while for higher redshifts, they
      are manually scaled to the expected 3.3\um PAH fluxes of high-z
      ULIRGs (see text for more details). The broadband estimated PAH
      fluxes are indicated by the difference between the
      spectro-photometry in the filter covering the $3.3$\um PAH
      feature and its local continuum calculated by the linear
      interpolation of its two neighboring filters. The red solid and
      dashed lines indicate the best fit line (shown at the top left
      in each panel) and its 1$\sigma$ uncertainty, respectively.}
    \label{fig:JWST_PAH_flux}
  \end{center}
\end{figure}

In Figure~\ref{fig:JWST_PAH_flux}, we show the 3.3\um PAH flux as
measured with our AKARI spectra against the 3.3\um color excess.  At
$z=0$, the PAH fluxes are the actual measurements from our spectra,
but for higher redshifts, at each redshift we arbitrarily scale all of
the 3.3\um PAH fluxes with a single scaling factor to the expected
values for high-z ULIRGs ($\gtrsim 10^{12}\,L_\odot$).  As discussed
above, the expected fluxes are calculated assuming that the 3.3\um PAH
emission on average contributes about $0.1\%$ to the total infrared
emission.  At $z=0$, F336W covers the PAH feature, and F277W and F444W
are used to estimate its local continuum. The spectro-photometrically
estimated $3.3$\um excesses correlate quite well with the directly
measured $3.3$\um PAH fluxes in the {\it AKARI} spectra. This suggests
that we are able to infer $3.3$\um PAH fluxes based on the {\it JWST}
broadband photometry for most LIRGs.  The dispersion along the x-axis
is likely caused by the PAH sub-peak, $\rm H_2O$ ice absorption, and
CO absorption features in the {\it AKARI} spectral range.  If these
are indeed the reasons, this dispersion is real, and a fundamental
limit to estimating the 3.3\um PAH flux via photometry.  Among the
redshift bins in Figure~\ref{fig:JWST_PAH_flux}, the dispersion around
the regression fit is smallest at $z=3.5$ and $z=0.6$, which provides
the most reliable estimate of 3.3\um PAH fluxes.  There are more
substantial outliers seen at redshifts of $1.2$, $2.5$, and $4.0$. At
$z=1.2$, the 3.3\um PAH feature is covered by F770W, which has
$\lambda/\Delta\lambda = 3.5$, the smallest among the MIRI
filters. This may cause a larger uncertainty when estimating 3.3\um
PAH fluxes. At $z=2.5$, the PAH falls at the blue edge of F1280W and
it is only partially covered by the filter. At $z=4.0$, the PAH is in
both F1800W and F1500W, which may cause a larger scatter.  Although
the dispersions are larger in some redshift bins, this correlation
holds up to $z=5$, to the reddest {\it JWST} coverage.  We summarize
the coefficients for the correlation,
$F_{\rm 3.3 \, PAH} = 10^{a\cdot\log{(Excess)} + b}$, at each redshift
in Table~\ref{tbl:JWST_phot}.  At $z=3.5$, which shows the tightest
correlation, the photometric excess of 1 corresponds to
$F_{\rm 3.3 \, PAH} = (4.1 ^{+1.42}_{-1.05}) \times 10^{-19} \, {\rm
  W\,m^{-2}}$,
while in the redshift bin $z=1.2$, which has one of the largest
dispersions, the estimated flux has a larger uncertainty,
$F_{\rm 3.3 \, PAH} = (4.9 ^{+2.02}_{-1.43}) \times 10^{-18} \, {\rm
  W\,m^{-2}}$.
With the excess calculated using three {\it JWST} broadband filters,
these relations offer an estimate of 3.3\um PAH fluxes, which, in
particular, could be useful for identifying strong sources of PAH
emission (starbursts) in high-redshift groups or clusters
\citep[e.g.,][]{Duc02,Sain08,Koya08}.

\begin{table*}[h]
  \begin{center}
    \caption{Best fit linear coefficients relating 3.3\um PAH flux to
      {\it JWST} filter color excess from $z=0.0$ to $z=5.0$}
    \begin{tabular}{ccccc}
      \hline\hline
      \multirow{2}{*}{Redshift} & \multirow{2}{*}{a} & \multirow{2}{*}{b} & \multicolumn{2}{c}{{\it JWST} filters} \\
                                &                    &                    & 3.3\um PAH & Local cont.    \\
      \hline
      0.0 & $0.64 \pm 0.23$ & $-15.87 \pm 0.15$ &    F356W   &  F277W, F444W  \\
      0.3 & $0.39 \pm 0.26$ & $-15.83 \pm 0.16$ &    F444W   &  F356W, F560W  \\
      0.6 & $0.78 \pm 0.25$ & $-17.17 \pm 0.17$ &    F560W   &  F444W, F770W  \\
      1.2 & $0.57 \pm 0.26$ & $-17.31 \pm 0.15$ &    F770W   &  F560W, F1000W \\
      2.0 & $0.62 \pm 0.25$ & $-17.96 \pm 0.12$ &   F1000W   &  F770W, F1280W \\
      2.5 & $0.49 \pm 0.22$ & $-18.15 \pm 0.11$ &   F1280W   & F1000W, F1500W \\
      3.5 & $0.75 \pm 0.20$ & $-18.39 \pm 0.13$ &   F1500W   & F1280W, F1800W \\
      4.0 & $0.41 \pm 0.22$ & $-18.49 \pm 0.16$ &   F1800W   & F1500W, F2100W \\
      5.0 & $0.64 \pm 0.22$ & $-18.71 \pm 0.14$ &   F2100W   & F1800W, F2550W \\
      \hline
    \end{tabular}
    \tablefoot{ The coefficients of the best fit lines in
      Figure~\ref{fig:JWST_PAH_flux}:
      $F_{\rm 3.3 \, PAH} = 10^{a\cdot\log{(Excess)} + b}$, where
      $Excess$ is a value on the x-axis, the flux difference measured
      in the {\it JWST} filters covering the 3.3\um PAH feature and
      its neighboring filters (see Section~\ref{ss:JWST_excess} for
      more details).  The {\it JWST} filters which cover the 3.3\um
      PAH feature and the filters used for estimating its local
      continuum at each redshift are also listed.}
    \label{tbl:JWST_phot}
  \end{center}
\end{table*}


\section{Summary and Conclusions}\label{sec:summary}

We present analyses of 145 local LIRGs of the GOALS galaxy sample with
available {\it AKARI} spectra, covering the $2.5-5$\um rest-frame
wavelength.  We combine the measurements taken from the {\it AKARI}
spectra, such as the 3.3\um PAH feature, the Br$\alpha$ emission line,
and the $2.5-5$\um continuum slope, with mid-infrared measurements
obtained from {\it Spitzer} spectra, such as the 6.2\um PAH feature,
to study the nature of the local LIRG population.  Based on our
findings, we also explore potential starburst/AGN diagnostics in the
{\it JWST} era.  We have reached the following results:

\begin{enumerate}

\item The relation between the 3.3\um and 6.2\um PAH EQWs shows that
  starburst-dominated galaxies, identified by 6.2\um PAH EQW
  $\geq 0.6$\um, cover a wide range of the 3.3\um PAH EQW from
  $0.05$\um to $0.16$\um. Despite the large spread in the 3.3\um PAH
  EQW compared with the 6.2\um PAH EQW, the 3.3\um PAH EQW is an
  effective diagnostic for selecting starburst-dominated galaxies in
  most cases.

\item We find that a significant fraction ($\gtrsim 1/3$) of the
  galaxies with $\rm PAH(3.3{\rm \mu m}) \, EQW < 0.04$\um, which
  would typically be considered as AGN-dominated sources, have high
  6.2\um PAH EQWs ($\geq 0.4$\um), suggesting that they are instead,
  starburst dominated.  These galaxies have {\it AKARI} continua as
  blue as those showing high EQWs in both 3.3\um and 6.2\um PAH
  features. While the median {\it Spitzer} spectra of these two types
  of sources are remarkably similar, their median {\it AKARI} spectra
  are significantly different.
  We suggest that they are in fact galaxies whose $2-5$\um continua
  are dominated by strong stellar emission rather than hot dust, which
  also reduces the 3.3\um PAH EQW to very low levels.  For this class
  of galaxy, measuring the near- to mid-infrared continuum slope is
  critical to properly assess the dominant power source.  The use of
  the 3.3\um PAH EQW alone is not adequate to select AGN-dominated
  galaxies in all cases.

\item We exploit the 6.2\um PAH EQW as well as our estimated
  bolometric AGN contribution provided by the {\it Spitzer}/IRS
  spectra to propose a revised starburst/AGN diagnostic diagram based
  on $2.5-5$\um spectroscopic data.
  We use the 3.3\um PAH EQW and the $2.5-5$\um continuum slope, as
  measured by the $F_\nu(4.3{\rm \mu m})/F_\nu(2.8{\rm \mu m})$ flux
  ratio, to establish that starburst-dominated sources have
  $\rm PAH(3.3{\rm \mu m}) \, EQW \geq 0.06$\um, while AGN-dominated
  sources have both $\rm PAH(3.3{\rm \mu m}) \, EQW < 0.06$\um and
  $F_\nu(4.3{\rm \mu m})/F_\nu(2.8{\rm \mu m}) \geq 1.0$.

\item A clear correlation is seen between $L_{{\rm Br}\alpha}$ and
  $L_{IR}$ regardless of the galactic energy source
  when the dust extinction correction is applied using the 9.7\um
  silicate absorption. The SFRs estimated from the extinction
  corrected Br$\alpha$ agree well with SFRs estimated from the total
  infrared luminosity.

\item Based on the combined 2MASS, {\it AKARI}, and {\it Spitzer} data
  for the individual galaxies, we have explored starburst/AGN
  diagnostics with the {\it JWST} filters.  Commonly used {\it
    Spitzer}/IRAC color selection diagrams of \cite{Lacy04} and
  \cite{Donl12} are adapted for the {\it JWST} filters.  While they
  perform well up to $z \sim 0.3$, at higher redshifts,
  starburst-dominated galaxies start to contaminate the AGN selection
  of \cite{Lacy04}, while the boundaries of \cite{Donl12} work well at
  least up to $z=1.5$.

\item Although a clean selection of AGN-dominated sources is
  difficult, one of the simplest starburst selections which can be
  used up to $z \sim 5$ with {\it JWST} is to measure the rest-frame
  $2-5$\um continuum slope using the broadband filters to exclude
  galaxies with hot dust emission. The rest-frame $2-5$\um continuum
  color cut at $< 1.5$ (e.g., F1280W/F560 for $z \sim 2.0$)
  successfully retrieves $70-80\%$ of the starburst-dominated
  galaxies.

\item Artificially redshifting the combined 2MASS, {\it AKARI}, and
  {\it Spitzer} spectra from $z=0$ to $5$, we measure the excess of
  3.3\um PAH emission in a {\it JWST} filter using its neighboring
  filters to remove the local continuum. This excess correlates well
  with the 3.3\um PAH fluxes measured directly in the {\it AKARI}
  spectra up to $z=5$. In particular, we find that the most reliable
  estimate is at $z = 3.5$ and $z = 0.6$.  This correlation can be
  useful for identifying obscured star formation in high-$z$ galaxies
  with {\it JWST} photometry.

\end{enumerate}

\begin{acknowledgements} 

  The authors would like to thank the referee whose constructive
  comments helped improving the manuscript.  HI thanks Grant-in-Aid
  for Japan Society for the Promotion of Science (JSPS) Fellows
  (21-969) and JSPS Excellent Young Researchers Overseas Visit Program
  for supporting this work at the Spitzer Science Center, California
  Institute of Technology, USA.  YO acknowledges support from Ministry
  of Science and Technology (MOST) of Taiwan 106-2112-M-001-008-.
  This research is based on observations with AKARI, a JAXA project
  with the participation of ESA.  The authors appreciate the
  opportunity to present this work at the 4th AKARI international
  conference which helped to enrich this work \citep{Inam18proc}.
  The Spitzer Space Telescope is operated by the Jet Propulsion
  Laboratory, California Institute of Technology, under NASA contract
  1407.  This research has made use of the NASA/IPAC Extragalactic
  Database (NED) and the Infrared Science Archive (IRSA) which are
  operated by the Jet Propulsion Laboratory, California Institute of
  Technology, under contract with the National Aeronautics and Space
  Administration.

\end{acknowledgements}

\bibliography{bib}

\clearpage
\begin{appendix}

\section{The redshifted median SEDs of starburst and AGN-dominated
  sources}\label{app:z_sed}

\begin{figure*}
  \begin{center}
    \includegraphics[width=0.65\textwidth,trim=60 -350 -50 -10,angle=90,origin=c]
    {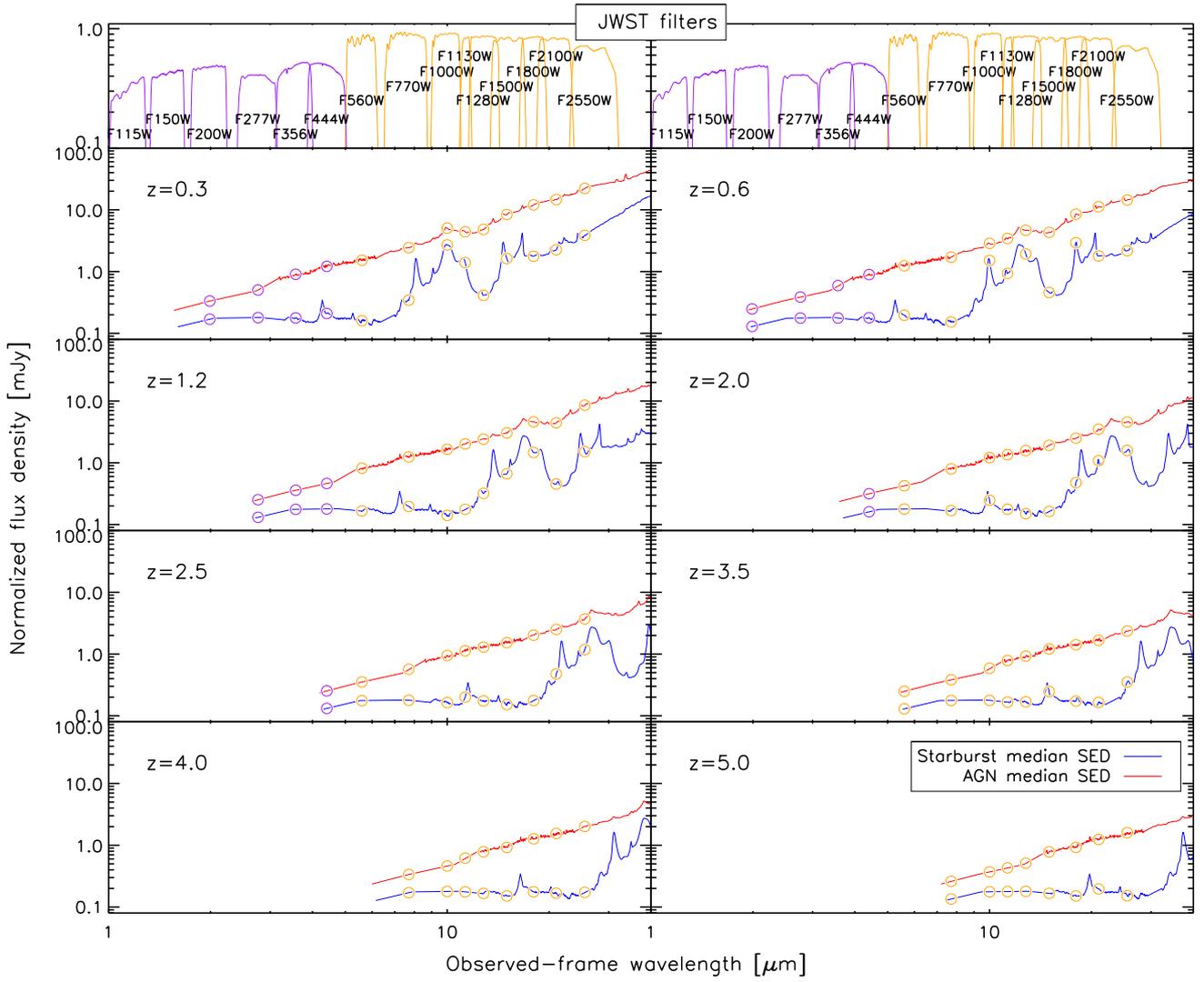}
    \caption{ The same plots as Figure~\ref{fig:JWST_SED}, but the
      SEDs are redshifted from $z=0.3$ to $5.0$ as indicated in each
      panel. The orange circles on each SED indicate the locations of
      the effective wavelengths of the {\it JWST} filters shown in the
      upper most panels. }
    \label{fig:JWST_SED_z}
  \end{center}
\end{figure*}

In Figure~\ref{fig:JWST_SED_z}, we show the redshifted median SEDs of
starburst- and AGN-dominated sources, which are presented in
Figure~\ref{fig:JWST_SED}. The redshift bins match the ones used in
Figures~\ref{fig:JWST_AGN_z} and \ref{fig:JWST_PAH_flux}.

\clearpage

\section{The {\it AKARI} Spectra}\label{app:akari_spec}

\begin{figure*}
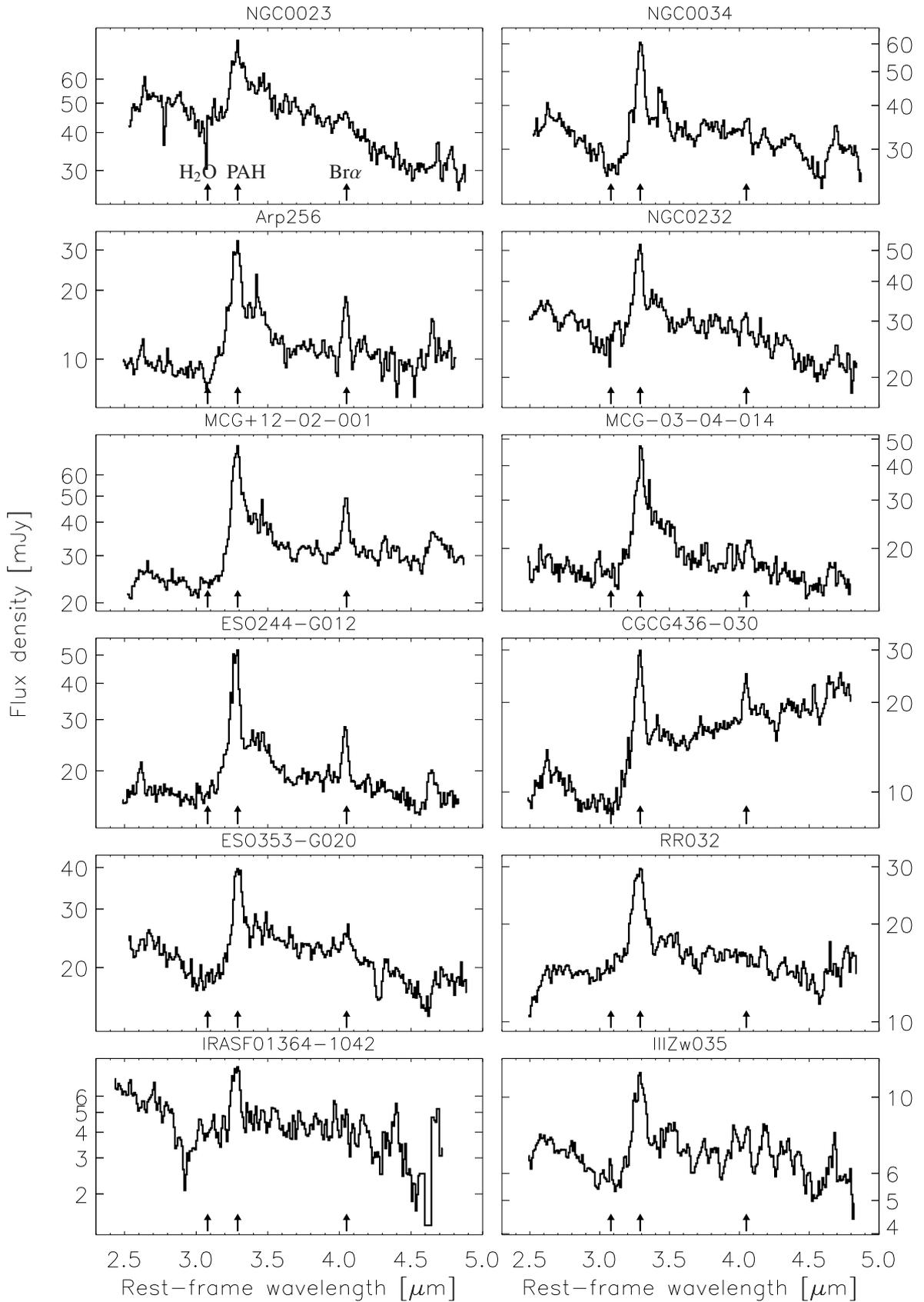

  \begin{center}
    \begin{overpic}
      {figures/20171201_all_spec_AperSL.ps_pages0}
      \put(16.0,85){${\rm H_2O}$}
      \put(19.5,85){PAH}
      \put(27,85){Br$\alpha$}
  \end{overpic}
  \caption{ The {\it AKARI} $2.5-5$\um spectra of the GOALS sample in
    the order of their R.A. and Dec coordinates. The arrows indicate the
    wavelengths of the $3.1$\um water ice absorption,
    3.3\,\um PAH emission, and $4.05$\um Br$\alpha$ emission
    features.}
    \label{fig:akari_spec}
  \end{center}
\end{figure*}

All of the {\it AKARI} spectra used in this paper are shown in
Figure~\ref{fig:akari_spec}.  The line fluxes and EQWs of 3.3\um PAH
and Br$\alpha$ and the $F_\nu(4.3{\rm \mu m})/F_\nu(2.8{\rm \mu m})$
color are reported in Table~\ref{tbl:fluxes}.

\begin{figure*}
  \addtocounter{subfigure}{1}
  \begin{center}
    \subfigure{
      \begin{overpic}
      {figures/20171201_all_spec_AperSL.ps_pages1}
      \put(16.,85){${\rm H_2O}$}
      \put(19.5,85){PAH}
      \put(27.5,85){Br$\alpha$}
    \end{overpic}
    }
  \end{center}
    Figure~\thefigure ~{\it Continued.}
\end{figure*}

\begin{figure*}
  \addtocounter{subfigure}{1}
  \begin{center}
    \subfigure{
      \begin{overpic}
        {figures/20171201_all_spec_AperSL.ps_pages2}
        \put(45,85){${\rm H_2O}$}
        \put(48.5,85){PAH}
        \put(56.5,85){Br$\alpha$}
      \end{overpic}
    }
  \end{center}
    Figure~\thefigure ~{\it Continued.}
\end{figure*}

\begin{figure*}
  \addtocounter{subfigure}{1}
  \begin{center}
    \subfigure{
      \begin{overpic}
        {figures/20171201_all_spec_AperSL.ps_pages3}
        \put(45.5,85){${\rm H_2O}$}
        \put(49,85){PAH}
        \put(56.5,85){Br$\alpha$}
      \end{overpic}
    }
  \end{center}
    Figure~\thefigure ~{\it Continued.}
\end{figure*}

\begin{figure*}
  \addtocounter{subfigure}{1}
  \begin{center}
    \subfigure{
      \begin{overpic}
        {figures/20171201_all_spec_AperSL.ps_pages4}
        \put(46,85){${\rm H_2O}$}
        \put(49.0,85){PAH}
        \put(56.5,85){Br$\alpha$}
      \end{overpic}
    }
  \end{center}
    Figure~\thefigure ~{\it Continued.}
\end{figure*}

\begin{figure*}
  \addtocounter{subfigure}{1}
  \begin{center}
    \subfigure{
      \begin{overpic}
        {figures/20171201_all_spec_AperSL.ps_pages5}
        \put(16.2,85){${\rm H_2O}$}
        \put(19.5,85){PAH}
        \put(27,85){Br$\alpha$}
      \end{overpic}
    }
  \end{center}
    Figure~\thefigure ~{\it Continued.}
\end{figure*}

\begin{figure*}
  \addtocounter{subfigure}{1}
  \begin{center}
    \subfigure{
      \begin{overpic}
      {figures/20171201_all_spec_AperSL.ps_pages6}
      \put(45.7,85){${\rm H_2O}$}
      \put(49.0,85){PAH}
      \put(56,85){Br$\alpha$}
    \end{overpic}
    }
  \end{center}
    Figure~\thefigure ~{\it Continued.}
\end{figure*}

\begin{figure*}
  \addtocounter{subfigure}{1}
  \begin{center}
    \subfigure{
      \begin{overpic}
        {figures/20171201_all_spec_AperSL.ps_pages7}
        \put(46.0,85){${\rm H_2O}$}
        \put(49.2,85){PAH}
        \put(56.5,85){Br$\alpha$}
      \end{overpic}
    }
  \end{center}
    Figure~\thefigure ~{\it Continued.}
\end{figure*}

\begin{figure*}
  \addtocounter{subfigure}{1}
  \begin{center}
    \subfigure{
      \begin{overpic}
        {figures/20171201_all_spec_AperSL.ps_pages8}
        \put(45.5,85){${\rm H_2O}$}
        \put(49.0,85){PAH}
        \put(56.5,85){Br$\alpha$}
      \end{overpic}
    }
  \end{center}
    Figure~\thefigure ~{\it Continued.}
\end{figure*}

\begin{figure*}
  \addtocounter{subfigure}{1}
  \begin{center}
    \subfigure{
      \begin{overpic}
        {figures/20171201_all_spec_AperSL.ps_pages9}
        \put(16.0,85){${\rm H_2O}$}
        \put(19.5,85){PAH}
        \put(27,85){Br$\alpha$}
      \end{overpic}
    }
  \end{center}
    Figure~\thefigure ~{\it Continued.}
\end{figure*}

\begin{figure*}
  \addtocounter{subfigure}{1}
  \begin{center}
    \subfigure{
      \begin{overpic}
        {figures/20171201_all_spec_AperSL.ps_pages10}
        \put(16.0,85){${\rm H_2O}$}
        \put(19.5,85){PAH}
        \put(27.0,85){Br$\alpha$}
      \end{overpic}
    }
  \end{center}
    Figure~\thefigure ~{\it Continued.}
\end{figure*}

\begin{figure*}
  \addtocounter{subfigure}{1}
  \begin{center}
    \subfigure{
      \begin{overpic}
        {figures/20171201_all_spec_AperSL.ps_pages11}
        \put(45.5,85){${\rm H_2O}$}
        \put(49,85){PAH}
        \put(56.5,85){Br$\alpha$}
      \end{overpic}
    }
  \end{center}
    Figure~\thefigure ~{\it Continued.}
\end{figure*}

\begin{figure*}[t]
  \addtocounter{subfigure}{1}
  \begin{center}
    \subfigure{
      \begin{overpic}
        [angle=0,scale=1.0,trim=0 500 0 0]{figures/20171201_all_spec_AperSL.ps_pages12}
        \put(22.5,13.3){${\rm H_2O}$}
        \put(27.5,13.3){PAH}
        \put(38.5,13.3){Br$\alpha$}
      \end{overpic}
    }
  \end{center}
    Figure~\thefigure ~{\it Continued.}
\end{figure*}

\end{appendix}

\end{document}